\documentstyle[iopconf1,psfig]{article}
\newcommand{\ea}{{\it et~al.\/} }
\newcommand{\dg}{\nobreak^\circ}

\newcommand{\ho}{\mbox{$\mbox{H}_0$} }
\newcommand{\simlt}{\mbox{$\stackrel{<}{_{\sim}}$} }
\newcommand{\simgt}{\mbox{$\stackrel{>}{_{\sim}}$} }

\newcommand{\kmspmpc}{\,\hbox{km}\,\hbox{s}^{-1}\,\hbox{Mpc}^{-1}}

\newcommand{\ie}{{\it i.e.\ }}
\newcommand{\eg}{{\it e.g.\ }}

\newcommand{\hz}{Harrison-Zel'dovich }

\newcommand{\qrms}{\mbox{$Q_{rms-ps}$} }
\newcommand{\rms}{{\em rms} }
\newcommand{\uk}{$\mu$K}
\newcommand{\be}{\begin{equation}}
\newcommand{\ee}{\end{equation}}
\newcommand{\cls}{\mbox{$C_l^{(S)}$}}
\newcommand{\clt}{\mbox{$C_l^{(T)}$}}
\newcommand{\ob}{\mbox{$\Omega_{b}$}}
\newcommand{\lbd}{{$\Lambda$}}
\newcommand{\omeg}{\mbox{$\Omega_{0}$}}
\newcommand{\olbd}{\mbox{$\Omega_{\Lambda}$}}
\newcommand{\dtrms}{\mbox{$(\Delta T/T)_{rms}$}}
\newcommand{\conv}{\mbox{$C_{conv}$}}

\newcommand{\cl}{\mbox{$C_{l}$}}
\newcommand{\ctheta}{\mbox{$C(\theta)$}}
\newcommand{\cthsig}{\mbox{$C(\theta,\sigma)$}}
\newcommand{\sigcos}{\mbox{$\sigma_{cos}^{2}$}}
\newcommand{\sigsam}{\mbox{$\sigma_{sam}^{2}$}}

\def \dtotrms{$\Delta T_{rms}/T$ }
\def \o0{$\Omega_{0}$}

\def\deg{\ifmmode^\circ _\cdot\else$^\circ _ \cdot$\fi }    
\def\degg{\ifmmode^\circ \else$^\circ $\fi } 
\def \dtotrms{$\Delta T_{rms}/T$ }
\def\arcs{\ifmmode {'' }\else $'' $\fi}     
\def\buildrel#1\over#2{\mathrel{\mathop{\null#2}\limits^{#1}}}
\def\mper{\ifmmode \buildrel m\over . \else $\buildrel m\over .$\fi}
\def\hper{\ifmmode \rlap.^{h}\else $\rlap{.}^h$\fi}
\def\sper{\ifmmode \rlap.^{s}\else $\rlap{.}^s$\fi}
\def\arcsper{\ifmmode \rlap.{' }\else $\rlap{.}' $\fi}
\def\arcmper{\ifmmode \rlap.{'' }\else $\rlap{.}'' $\fi}

\def\et{{\it et~al.~}}
\newcommand{\lta}{{\small\raisebox{-0.6ex}{$\,\stackrel
{\raisebox{-.2ex}{$\textstyle <$}}{\sim}\,$}}}
\newcommand{\gta}{{\small\raisebox{-0.6ex}{$\,\stackrel
{\raisebox{-.2ex}{$\textstyle >$}}{\sim}\,$}}}

\def\apj{ApJ}  
\def\mn{MNRAS}      

\begin{document}

\title{Constraints on the Cosmological Parameters using CMB observations}

\author{Gra\c{c}a Rocha\dag\footnote{E-mail:
graca@phys.ksu.edu},\ddag\footnote{E-mail:graca@astro.up.pt}}

\affil{\dag\ KSU, Kansas State University, Department of Physics,
116 Cardwell Hall,
Manhattan, Kansas 66506-2601, USA}

\affil{\ddag\ Centro de Astrof\'{\i}sica da Universidade do Porto, 
Rua das Estrelas s/n, 4100 Porto, Portugal.}

\beginabstract
This paper covers several techniques of intercomparison of Cosmic Microwave Background (CMB) anisotropy experiments and models of structure formation. It presents the constraints on several cosmological parameters using current CMB observations.
\endabstract

\small
\section{Introduction}

In 1964 Penzias and Wilson \cite{penzias} discovered the Cosmic Microwave Background (CMB) radiation as an excess instrument noise with temperature $\sim$ 3K, coming from all directions in the sky.
Dicke and Peebles \ea (1965) \cite{dicke} interpreted this as the relic radiation from the Big Bang.
The discovery of this relic radiation in conjunction with Hubble's observation of the cosmic expansion in 1927 \cite{hubble}, the synthesis of the light elements, the evolution of radio source counts etc, provided strong support to the hot Big Bang model.

Recently the FIRAS instrument on the COBE satellite has confirmed that this radiation has a Planck spectrum with \cite{mather}:
\be
T_{0}=2.726 \pm 0.010 \rm{K}\ \rm{(95\%\ confidence )}.
\ee
In the standard Big Bang model thermalization of the CMB occurs at a 
redshift $z \sim 10^{6}$ at $T \geq 10^{7}$K, via double Compton scattering 
($e+\gamma \rightarrow e+2 \gamma$) and bremsstrahlung ($e+p \rightarrow e+p+ \gamma$) processes, resulting in a Planckian spectrum kept along its subsequent evolution.
So, the observed black body spectrum of the CMB radiation suggests its origin in the early universe; this is further supported by its high degree of isotropy.
The CMB photons were last scattered at a redshift $z \sim 1100$, when the 
expanding universe had cooled to temperatures $T \sim  3000 $K, sufficient for protons to capture the free electrons producing neutral atoms.
After this period of recombination the CMB photons free stream towards us from 
this `last scattering surface', providing direct information about this early epoch 300,000 years after the initial singularity.
Initial theoretical predictions suggested that structure formation in the universe like galaxies, clusters of galaxies, etc., implied fluctuations in the CMB, due to the presence of primordial seed structures at the time when the CMB radiation decouples from the matter distribution.
Consequently observations of CMB radiation anisotropies provide a means of testing stucture formation theories. On large angular scales the CMB fluctuations may probe the inflationary power spectrum and therefore the physics of the
first $10^{-35}$s.
The original detection paper by Penzias and Wilson 1965 \cite{penzias}, set an upper limit of $<10 \%$ on any anisotropy. 
The initial predictions indicated that galaxy formation would produce 
fluctuations in the CMB of the order of 1 part in $10^{2}$ \cite{sachswolfe}, or $10^{3}$ \cite{silk67,silk68}.

Since then work on the CMB anisotropy question has shown improvements in experimental sensitivity and in theoretical calculations with increasingly 
sophisticated estimations of the temperature fluctuations of the CMB.
For universes containing predominantly baryonic matter, a prediction of 
$\Delta T / T \sim 10^{-4}$ was obtained by several authors \cite{peeblesyu,doros,wilson}.
By 1980, the hypothesis of non-baryonic dark matter universes
\cite{bond80,doros80} was encouraged by the claims of the existence of a electron neutrino mass of $\sim$ 30eV \cite{lubi}. Although the neutrino as a candidate of dark matter failed to explain the structure formation \cite{kaiser83,white83,white84}, it still is an acceptable dark matter candidate \cite{steigman}.
At the same time inflationary scenarios paid particular attention to cold dark matter dominated universes \cite{blumenthal,peebles82,frenk90}, where cold means that the candidate particles have small thermal velocities (the free streaming length of such particles is much smaller than the size of any galaxy or cluster of stars).
Following improvements in the upper limits on smaller scales of 
$\Delta T / T \sim 10^{-4}$ \cite{usonwilk1,usonwilk2,usonwilk3}, new theoretical calculations predicted that for models with a cold dark matter component, the amplitude of the temperature fluctuations of the CMB could be reduced by an order of magnitude.

The first detection of large angular scale primordial CMB anisotropy came in 1992, reported by Smoot \ea \cite{smoot} using the COBE DMR experiment \cite{smoot90}.
This initial result consisted of statistical evidence for CMB anisotropy; the
 CMB maps were not sensitive enough to pin down actual CMB features on the sky.
In 1993 the MIT experiment \cite{meyer91,Ganga93} reported results at high frequency which confirmed the COBE detection.
Given the frequency range and the sky coverage between these two experiments 
this confirmed that the detected signal was indeed CMB anisotropy.
This was followed by low frequency confirmation in 1994, from the Tenerife experiment \cite{nature94}, which enabled the association of the
detected statistical anisotropy with actual CMB features on the sky.
Since the COBE detection, several other experiments have reported detections 
on intermediate angular scales (see Section~\ref{exper}), although  given their sky coverage and frequency range spanned, the possibility of foreground contamination is in some of the cases still under consideration.
The systematic effects affecting some of these observations are another cause of concern.
Recently new interesting detections have been reported by the Saskatoon \cite{sask} and the CAT \cite{cat} experiments, among others, which with the other observations cover angular scales ranging from $10 \dg$ to $0.25 \dg$.
Meanwhile questions such as the process of structure formation, the origin of the seed perturbations, the nature and quantity of the matter content of the universe, the values of the density parameter $\Omega$, of the Hubble constant \ho, and of other cosmological parameters remain still to be answered conclusively.
 
All these new exciting results and more sophisticated theoretical calculations motivated the work carried out by several teams in search of satisfactory answers to such questions, bearing in mind the limitations imposed by the status of the data.

\section{Essential concepts and mathematical tools}
\label{basic}

\subsection{Statistical description of the CMB field \label{stat}}

It is standard procedure to express the temperature fluctuations of the CMB on the celestial sphere using an expansion in spherical harmonics:

\begin{equation}
\Delta T/T(\alpha,\phi)=\sum_{l=2}^{\infty} \sum_{m=-l}^{l} a_{lm} Y_{lm}(\alpha,\phi) \label{eq:dt}
\end{equation}
where $(\alpha,\phi)$ are the polar coordinates of a point in the spherical surface.
We have ignored the dipole ($l=1$) which is largely dominated by the Doppler effect due to the motion of our Galaxy with respect to the last scattering surface. Inflationary models predict that the initial CMB perturbations follow a Gaussian distribution, and as the observed CMB field conserves the original distribution we can consider it as a 2-dimensional Gaussian random field on the sky. In this case the multipole moments $a_{lm}$ are stochastic variables with random phase, zero mean and variance given by $C_{l}=\langle |a_{lm}|^{2} \rangle $ which defines a power-spectrum.
The properties of a Gaussian field are fully specified by the \mbox{two-point} angular \mbox{auto-correlation} function (ACF), the expected average of the product of the temperature fluctuations in two directions in the sky separated by an angle $\theta$ :

\be
 C(\theta)=\left\langle \frac {\Delta T}{T} (\vec{n_{1}}) \frac{\Delta T}{T} (\vec{n_{2}}) \right\rangle
\label{eq:ctheta}
\ee

where $\vec{n_{1}}\cdot\vec{n_{2}}=\cos\theta$ and the angle brackets refer to an average over the whole sky.
Using Equation~(\ref{eq:dt}) the expression of the ACF becomes:
\be
C_{ran}(\theta)=\sum_{l,m} \sum_{l',m'} \langle a_{lm} a^{\ast}_{l'm'} \rangle Y_{lm}(\alpha,\phi) Y^{\ast}_{l'm'} (\alpha',\phi') \label {eq:ct}
\ee
For a temperature pattern that is statistically isotropic, the rotational symmetry implies:

\be
\langle a^{\ast}_{lm} a_{l'm'} \rangle = \delta_{l l'} \delta_{m m'} C_{l}. \label {eq:alm}
\ee
Using  the relation~(\ref{eq:alm}) and the addition theorem 
for spherical harmonics, one can express the random variable $C(\theta)$, 
as an expansion in Legendre polynomials \cite{cosvariance1}:
\begin{equation} 
C_{ran}(\theta)=\frac{1}{4\pi} \sum_{l\geq 2} Q_{l}^{2}(x) P_{l}(\cos\theta), \label{eq:cthetavar}
\end{equation}
 where
\be
 Q_{l}^{2}(x)=\sum_{m=-l}^{l} |a_{lm}|^{2},
\ee
 with a $\chi_{2l+1}^{2}$ distribution and
\be
\langle Q_{l}^{2}(x) \rangle =(2l+1) C_{l},
\ee
and variance
\be
 \mbox{var}(Q_{l}^{2})=2 \langle Q_{l}^{2} \rangle ^{2}/(2l+1). \label{eq:cos}
\ee
The relationship between the angular scale $\theta$ and the multipole $l$ is given by $l\simeq 2/\theta $, with $\theta$ in radians.
The coefficients $C_{l}$ constitute the angular power spectrum of the temperature fluctuations and, as we will see, play an important role in the comparison of CMB experiments with theoretical models of structure formation. 
As shown in Equation~(\ref{eq:cos}) the uncertainty scales as $(l+ 1/2)^{-1/2}$ which explains the fact that smaller angular scale experiments have smaller cosmic variances (see also Section~\ref{4}), since they probe larger multipoles $l$.
For large scale experiments this is an unavoidable uncertainty and in particular it limits the accuracy of the normalization of the power spectrum. 
The expectation value of the temperature correlation function, i.e., the ensemble average, is given by:
\begin{equation}
C(\theta)=\frac{1}{4\pi}\sum_{l}(2l+1)C_{l}P_{l}(\cos\theta). \label{eq:cthetaens}
\end{equation}

In this statistical description of the CMB fluctuations, our universe is considered as a realization of the theoretical field, and therefore the observed ACF represents an estimation of the ACF of the field from a finite sample. The uncertainty of these estimations constitutes an intrinsic limitation imposed by the statistical nature of the theory and is independent of the instrumental sensitivity.


\subsection{($\Delta T/T)_{rms}$ \label{2}}

One of the observables in CMB experiments is the rms of temperature fluctuations. 
This quantity may then be compared with the expected temperature fluctuation for a given experiment and model. For a rigorous comparison the theoretical uncertainty must also be considered (see \cite{thesis}).

The expected temperature fluctuation, $(\Delta T/T)_{rms}$, may be computed directly from the two-point angular correlation function, \ctheta, or from the angular power spectrum, \cl.
The former may be used if the theory provides an analytical expression for the angular correlation function (e.g., a fitted parameterized expression) or, e.g., a parameterized expression for the transfer function of the temperature fluctuations (for transfer function see e.g \cite{thesis}). 

The expression of $(\Delta T/T)_{rms}$ depends on the experimental configuration, e.g., beam size, switching pattern, etc. Most CMB experiments use beam-switching techniques, using e.g. single, double differences etc., i.e., double or triple beams respectively. 
A two-beam experiment measures the temperature difference between beams separated by an angle $\beta$ on the sky 
\be
\Delta T/T=(T_{1}-T_{2})/T,
\ee
and its expected variance is written in terms of the temperature autocorrelation function :
\be
\left(\frac{\Delta T}{T}\right)_{rms}^{2}=\left\langle \left(\frac{T_{1}-T_{2}}{T} \right)^{2}\right\rangle=2(C(0)-C(\beta)).
\ee
A three-beam experiment measures the difference between a field point, $T_{f}$, and the mean value of the temperature in two directions which are separated from the field point by an angle $\beta$
\be
\Delta T/T=T_{f}-\textstyle{\frac{1}{2}}(T_{1}+T_{2}),
\ee
and its expected variance is given by:
\be
\left(\frac{\Delta T}{T}\right)_{rms}^{2}=
\left\langle \left(\frac{T_{f}-\frac{1}{2}(T_{1}+T_{2})}{T}\right)^{2}\right\rangle
=\textstyle{\frac{3}{2}}C(0)-2C(\beta)+\textstyle{\frac{1}{2}}C(2\beta)
\ee 
A four-beam experiment measures e.g.:
\be
\Delta T_{j}/T=-\textstyle{\frac{1}{4}}T_{j}+\textstyle{\frac{3}{4}}T_{j+1}-\textstyle{\frac{3}{4}}T_{j+2}
+\textstyle{\frac{1}{4}}T_{j+3}. 
\ee
Where $\angle\left(T_{j},T_{j+1}\right)=\beta$, the beamthrow, and its expected variance is expressed by: 
\be
\left(\frac{\Delta T}{T}\right)_{rms}^{2}=\textstyle{\frac{5}{4}}C(0)-
\textstyle{\frac{15}{8}}C(\beta)+\textstyle{\frac{3}{4}}
C(2\beta)-\textstyle{\frac{1}{8}}C(3\beta).
\ee
All these expressions are true for an ideal experiment with an infinitely narrow beam; for a real observation it is necessary to consider the beam smearing due to the finite resolution of the antenna. This is usually taken into account by convolving the radiation intensity with a beam approximated by a Gaussian $G(\theta)$:
\be
G(\theta)=\frac{1}{2\pi{\sigma}^{2}} \exp\left(-\frac{{\theta}^{2}}{2{\sigma}^{2}}\right), \label{eq:go}
\ee
where $\sigma$ is the dispersion of the Gaussian. The correlation function of the convolved radiation field is given by \cite{be87}:
\begin{equation}
C(\theta,\sigma)\simeq \frac{1}{2{\sigma}^{2}}\int_{0}^{\infty}C(\theta')e^{-\frac
{{\theta}^{2}+{\theta'}^{2}}{4{\sigma}^{2}}}I_{0}\left(\frac{\theta\theta'}{2{\sigma}^{2}}\right)\theta'd\theta',  \label{equm}
\end{equation}
 where $I_{0}$ is the modified Bessel function.
So for a single difference one has:
\begin{equation}
\left(\frac{\Delta T}{T}\right)_{rms}^{2}=2(C(0,\sigma)-C(\beta,\sigma))  \label{eqdois}
\end{equation}
and for a double difference:
\begin{equation}
\left(\frac{\Delta T}{T}\right)_{rms}^{2}=\textstyle{\frac{3}{2}}C(0,\sigma)-2C(\beta,\sigma)+\textstyle{\frac{1}{2}}C(2\beta,\sigma),  \label{eqtres}
\end{equation}
where $\beta$ is the beamthrow. 
For a four-beam experiment:
\be
\left(\frac{\Delta T}{T}\right)_{rms}^{2}=
\textstyle{\frac{5}{4}}C(0, \sigma)-\textstyle{\frac{15}{8}}C(\beta, \sigma)+\textstyle{\frac{3}{4}}
C(2\beta, \sigma)-\textstyle{\frac{1}{8}}C(3\beta, \sigma)  \label{equatro}
\ee
One is able to compute $(\frac{\Delta T}{T})_{rms}^{2}$ using Equations (\ref{equm}) and (\ref{eqdois}), (\ref{eqtres}) or (\ref{equatro}) according to the experimental configuration.

The use of the angular power spectrum has in the last few years become the most successful representation and methodology for treatment of CMB anisotropies. 
The use of this approach is possible because the proponents of several models of structure formation are now making predictions of the power spectrum of the fluctuations. This approach uses the expression of \ctheta\ as in Equation~(\ref{eq:cthetaens}), which considers ensemble-averaged values.
As an example of this approach consider the particular case of a single Gaussian beam (the general case is addressed in Section~\ref{3}), for comparison with the description given previously.
In this case the smearing effect due to finite resolution of the antenna is taken into account in terms of a function ${\cal W}_{l}$ which affects the spherical harmonic expansion of the temperature fluctuations: 
\be
\frac{\Delta T}{T}(\alpha,\phi,\sigma_{b})=\sum_{l,m}a_{lm}Y_{lm}(\alpha,\phi){\cal W}_{l}(\sigma_{b}). \label{eq:dttconv}
\ee
This is translated in terms of a window function $W_{l}$ acting on the \cl\ expansion of the correlation function:
\begin{equation}
 C(\theta,\sigma)=\frac{1}{4\pi}\sum_{l}(2l+1)C_{l}P_{l}(\cos\theta)W_{l}(\sigma),  \label{eqcinco}
\end{equation}
 where \cite{cosvariance1} 
\be
{\cal W}_{l}(\sigma)=e^{-(2l+1)^{2}\sigma^{2}/8}
\ee
and 
\be
W_{l}(\sigma)=e^{-(2l+1)^{2}\sigma^{2}/4} \label{eq:wl}
\ee
which is usually approximated by
\be
W_{l}={\exp}(-l(l+1)\sigma^{2}), \label{eq:wlg} 
\ee

where $\sigma={\rm FWHM} /2 \sqrt{2\ln2}$, with FWHM standing for full width at half maximum of the beam.
Equation~\ref{eqcinco} is a particular case, for a single Gaussian beam, of a more general expression which is given in Section~\ref{3}. 
Once given the $C_{l}$'s one can compute the correlation using Equation~(\ref{eq:cthetaens}) or (\ref{eqcinco}) and $\Delta T/T$ from Equations~(\ref{eqdois}), (\ref{eqtres}) or (\ref{equatro}), for the 2, 3 and 4-beam  experiments respectively. 
Some experiments using switching techniques do not use a square wave chop; however, it may be used as an approximation. 
If one wants to be more precise one should use the appropriate beam on the sky. It is standard procedure to describe the details of the observation in terms of the window function, $W_{l}$, which is then used in the \cl\ expansion of \dtrms\ in order to compute directly the expected temperature fluctuations: this is addressed in the following Section (Section~\ref{3}).
 


\subsection{Window functions \label{3}}

A real CMB experiment is only sensitive to a given range in angular scales with
 a response to each multipole that can be characterized by the window function, $W_{l}$. 
For example, the beam smearing due to finite resolution results in the convolution of the radiation intensity with a beam approximated by a Gaussian $G(\theta)$. In the multipole space this will correspond to a multiplication of the coefficients of the Legendre polynomials by the 
corresponding window function, as given by Equation~(\ref{eqcinco}) and (\ref{eq:wl}). 
As already mentioned 
the intrinsic auto-correlations function $C_{int}$ is transformed by convolution with the autocorrelated beam pattern of the telescope and is given by $C_{conv}(\theta)=C_{int}(\theta) \star C_{beam}(\theta)$, 
where \conv\ is the autocorrelation function of the convolved 
radiation intensity, incorporating all the effects of the observational strategy e.g. the finite resolution of the antena, the switching pattern etc., $C_{beam}$ is the instrumental ACF defined by the configuration of the telescope and $\star$ refers to convolution. The beam pattern of the telescope depends on the experimental strategy and for example is given by a triple beam pattern for double differences experiments.
For a real observation Equation~(\ref{eq:cthetaens}) becomes
\be
\conv (\theta)=\frac{1}{4\pi}\sum_{l}(2l+1)C_{l}P_{l}(\cos\theta)W_{l}. \label{eq:cthetaconv}
\ee
From now on $C({\theta})$ will represent the autocorrelation function of the intrinsic radiation intensity.
The expression for the rms temperature fluctuations becomes
\be
\left(\frac{\Delta T}{T}\right)_{rms}^{2}=
\frac{1}{4\pi}\sum_{l}(2l+1)C_{l} W_{l},  \label{eq:dtconv}
\ee
where we are considering ensemble average values.
For a Gaussian beam pattern the window function describing the finite beam 
resolution is given by 
Equation~(\ref{eq:wlg})
\cite{cosvariance1} 
The multipole, $l_{s}$, corresponding to the dispersion of the beam, $\sigma$, 
is given by $l_{s}+1/2=1/(2 \sin(\sigma /2))$ \cite{bond1}.
This function determines the maximum resolution or the minimum angular scale to which the instrument is sensitive. It corresponds to a high $l$ cut off which is controlled by the beamwidth of the experiment.
\newline
The window function is given by, for a 2-beam experiment:
\begin{equation}
W_{l}={\exp}(-l(l+1)\sigma^{2})\left[2(1-P_{l}(\cos(\beta)))\right].
 \label{eq:2}
\end{equation}
for a 3-beam experiment:
\begin{equation}
W_{l}={\exp}(-l(l+1)\sigma^{2})\left[\textstyle{\frac{1}{2}}(3-4P_{l}(\cos(\beta))+P_{l}(\cos(2\beta)))\right]. \label{eq:3}
\end{equation}
and for a 4-beam experiment:
\be
W_{l}={\exp}(-l(l+1)\sigma^{2})\left[\textstyle{\frac{5}{4}}-\textstyle{\frac{15}{8}}P_{l}(\cos(\beta))+ \textstyle{\frac{3}{4}}P_{l}(\cos(2\beta))
-\textstyle{\frac{1}{8}}P_{l}(\cos(3 \beta)) \right]
\label{eq:4}
\ee
where $\beta$ is the beamthrow.
These expressions may be easily derived from Equations~(\ref{eq:dtconv}), (\ref{eqdois}),(\ref{eqtres}), and (\ref{equatro}).
The differencing techniques introduce a low $l$ cutoff which is determined by the beamthrow. As mentioned above there are experiments that do not use a `square wave chop', in which case the window function needs to be derived separately, but in most of these cases the observing strategy may still be approximated as a single or double difference switching. For example, the experiments at the South-Pole (SP91 \cite{gaier92,schuster93}, SP94 \cite{spole}), MAX \cite{Clapp94,devlin,max} and Saskatoon \cite{wollack93} use a sine wave chop with a weighting of $\pm$ 1 for the South-Pole, and the first and second harmonic of the chop frequency for MAX and Saskatoon respectively \cite{white,whitewindow1,whitewindow2}. All these cases may be incorporated in a generic expression of the window function, according to White and Srednicki \cite{whitewindow2}.

The Tenerife experiment \cite{nature94}, MSAM3 \cite{msam2} and OVRO \cite{Myers93,readhead89} are examples of a 3-beam experiment, MSAM2 \cite{msam1} is an example of a 2-beam experiment while PYTHON \cite{dragovan94,python} is an example of a 4-beam experiment.
Other small angle experiments use different lock-in functions (ie weighting functions which express the weight given to each point of the beam trajectory contributing for a given point of the scan, for the calculation of temperature, and the overall normalization), e.g.
 South-Pole (SP91) \cite{schuster93,gaier92} and ARGO \cite{argo} experiments use a `square-wave lock-in', the MAX experiment \cite{Clapp94,devlin,max} uses a `sine wave lock-in', and the Saskatoon experiment \cite{wollack93} uses a double angle ``cosine lock-in'' (for details see \cite{whitewindow2,thesis}).   
For the South-Pole experiment (SP91) the window function is given by:
\be
W_{l}=4 \frac{4 \pi}{2l+1} B_{l}^{2}(\sigma)
\sum_{m=-l}^{l} |Y_{lm}(\theta,\phi)|^{2} H_{0}^{2}(m \alpha) \label{eq:sp}
\ee
where $H_{0}(x)$ is the Struve function and $B_{l}^{2}$ is given in Equation (\ref{eq:wl}) or (\ref{eq:wlg}). 
For the MAX experiment, considering $\alpha_{0}=0.65^{\circ}$ and $\sigma=0.425 \times 0.5^{\circ}$, one has $k=\sqrt{1.13} \times \pi=3.34$, the window function is given by:
\be
W_{l}=k^{2} \frac{4 \pi}{2l+1} B_{l}^{2}(\sigma) \sum_{m=-l}^{l} 
|Y_{lm}(\theta,\phi)|^{2} J_{1}^{2}(|m \alpha_{0}|). \label{eq:max}
\ee 
where $J_{1}$ is the Bessel function. 
For Saskatoon with $\alpha_{0}=2.45^{\circ}$ and $\sigma=0.425 \times 1.44^{\circ}$ one has $k=\sqrt{1.74} \times \pi=4.14$, the window function is given by:
\be
W_{l}=k^{2} \frac{4 \pi}{2l+1} B_{l}^{2} \sum_{m=-l}^{l}
|Y_{lm}(\theta,\phi)|^{2} J_{2}^{2}(m \alpha_{0}). \label{eq:saska}
\ee 

Other experiments use interferometry techniques (e.g., CAT \cite{cat}), where the window function is the UV-sampling convolved with the Fourier Transform of the primary beam and is usually given by the observers (see also Section~\ref{comp1}, \cite{thesis}). An easy way to parameterise the window functions is by using three numbers which represent the multipole at which the instrument achieves its maximum sensitivity, and the multipoles at which the sensitivity decreases by a factor 2 from the peak, these are given in Table~\ref{tabdat} of Section~\ref{comp1} and Table~\ref{datup2} of Section~\ref{comp2}. Shown in 
Fig~\ref{fig:windows} of Section~\ref{comp1} are the window functions for the various observational configurations of several experiments. 
According to White and Srednicki \cite{whitewindow2,dodsteb94}, the approximation of the South-Pole experiment by a 2-beam, gives a difference with respect to the exact window function of the order 20\%; while for MAX it is of the order of 10\%.


\subsection{Statistical uncertainties \label{4}}

The theories predict only the statistics of the temperature fluctuations and ensemble average values, but not their value in our universe. The \cthsig\ is a random variable with ensemble average value given by Equation~(\ref{eqcinco}). It may be expressed as:
\begin{equation} 
C_{ran}(\theta,\sigma)=\frac{1}{4 \pi} \sum_{l \geq 2} Q_{l}^{2}(x)
 P_{l}(\cos \theta)W_{l}(\sigma), \label{equatorze}
\end{equation}
where all the quantities are as defined in Section~\ref{stat} and \ref{3}. The cosmic variance, \sigcos, is defined in terms of var$(Q_{l}^{2})$ and represents an intrinsic limitation of the statistical description of the CMB fluctuations. As already mentioned in Section~\ref{stat} this intrinsic cosmic variance is different for the different angular scales depending on the number of independent samples in the sky on those angular scales. The variance is therefore larger for large angular scales. Several authors have considered how this uncertainty affects the estimation of the ACF. Cay\'{o}n et al. (1991) \cite{Cayon91} obtained a semi-analytic expression for the probability density function, pdf, of the temperature autocorrelation, ACF.
Scaramela \& Vittorio (1990) \cite{cosvariance2} used the multipole expansion of the ACF and computed their distribution by Monte Carlo simulations. Alternatively Bond \& Efstathiou (1987) \cite{be87,cdm2} derived an analytic expression to compute \sigcos, which is expressed in terms of the \cl: 
\begin{equation}
{\rm var}(C(\theta))=\frac{1}{2 \pi^{2}}  \sum_{l \geq 2} (2l+1) C_{l}^{2} P_{l}^{2} 
\end{equation}
Or in terms of the window function, $W_{l}$, of a given experiment:
\begin{equation}
{\rm var}(C_{conv}(0))=\frac{1}{2 \pi^{2}}  \sum_{l \geq 2} (2l+1) C_{l}^{2} W_{l}^{2}. \label{eqdezasseis}
\end{equation}
Equation~(\ref{eqdezasseis}) could be extended to the cases where the window function has an $m$-dependence (White et al.).
Small angular scale experiments have negligible cosmic variance, since in this case, ergodicity applies and ensemble average is well represented by the angular average over the sky.  The problem is that if the observations sample only part of the sky, the number of independent samples in the sky at these angular scales will be reduced, this translates into an enhancement of the cosmic variance, called sample variance. Scott et al (1994) \cite{samvariance} derived an approximate relation between the sample variance (\sigsam) and the cosmic variance:

\begin{equation}
\sigma_{sam}^{2} \simeq \left(\frac{4 \pi}{A} \right)\sigma_{cos}^{2}  \label{eqdezassete}
\end{equation}
where $A$ is the solid angle covered by the experiment.
The above methods are applied to experiments considering, e.g., their scan strategy, appropriate solid angle, etc. The sample variance may be obtained by
computing the cosmic variance via 
e.g. (\ref{eqdezasseis}), or Monte Carlo simulations and then using the approximate relation given in Equation~(\ref{eqdezassete}).
We used a different approach, producing Monte Carlo simulations of the temperature fluctuations according to the specific characteristics of each experiment, such as the beamwidth, the sky coverage, the switching pattern, the beamthrow, etc. Proceeding in this way one computes directly the overall theoretical uncertainty for a given experiment. This method was of particular interest to us because we were concerned with the confrontation of frequentist with Bayesian techniques of data analysis. These simulations follow different strategies according to angular scale and experimental configuration. This is described in \cite{thesis}.
This technique can also be applied to test the reliability of a combination of data sets \cite{Ratra99b}.



\subsection{Angular power spectrum \label{5}}

Having discussed the effects of particular experimental arrangements we now discuss the form of the power spectrum itself. The description of the CMB anisotropy in terms of the angular power spectrum, \cl, has proved to be an invaluable method and has become a standard procedure for treatment of the temperature fluctuations of the CMB radiation. The competing models for the origin and evolution of structure, for example \cite{be87,hu} predict the shape and amplitude of the CMB power spectrum and its Fourier equivalent, the autocorrelation function $C(\theta)$.

\subsubsection{Power spectrum normalization}

\begin{itemize}
\item
The galaxy clustering normalization: $\sigma_{8}$
\end{itemize} 
The quantity, $\sigma _8$, is 
the dispersion of the density field in a sphere of radius $8 h^{-1}$Mpc. 
The observed quantity is the variance of counts of galaxies in spheres of the size $8 h^{-1}$Mpc, $\sigma_{8}(galaxies)$ which is related with the corresponding variance but in terms of mass, $\sigma_{8}(mass)$ via the bias parameter, $b$: 
\be
b^{2}=\frac{\sigma_{8}^{2}(galaxies)}{\sigma_{8}^{2}(mass)}.
\ee 
It is an observational result that $\sigma_{8}(galaxies)\simeq 1$ \cite{Holtzmann89}, this being so, one has: $b\simeq 1/\sigma_{8}(mass)$. 
Previous to the COBE detection the most common normalization of models used was based  on the value of $\sigma _8$, derived from galaxy clustering assuming a bias $b=1$ \ie that light traces mass \cite{kaiser84,davis85}. If in fact galaxies are more highly clustered than matter \cite{davis85,bardeen86}, the amplitude of the initial matter perturbations (and hence \dtotrms) necessary to produce the observed clustering is reduced by the factor $b$. The power on large scales depends on this normalization and therefore depends on the Hubble constant, $h$, ($h=\ho / 100 \kmspmpc$). It also depends on $\omeg h^{2}$, the quantity determining the deflection point of the power spectrum, where \omeg\ is the ratio of the total density of the universe to the critical density, for simplicty hereafter referred as to the total density of the universe.
The relevant quantity related with this normalization is the amplitude of baryonic fluctuations necessary at recombination in order to obtain the fluctuations we see today. This amplitude lowers if the growth factor increases consequently lowering  the normalization of the temperature fluctuations. This effect may be obtained, e.g., varying the value of the total density of the universe, \omeg\, or, e.g., considering non-baryonic dark matter. For example the lower the value of \omeg\ the lower the fluctuation growth therefore the larger the normalization of $\Delta T/T$.
The non-baryonic dark matter increases the growth of fluctuations, for instance in cold dark matter (CDM) models the CDM perturbations have an extra growth between the epoch of matter-radiation equality and the end of recombination $\propto \omeg h^{2}$ as compared to the baryonic perturbations. When baryons decouple at the end of recombination they fall into the potential wells of the non-baryonic dark matter and increase their amplitude to match the CDM perturbations.
A flat model with non-zero cosmological constant, \lbd, has a reduced growth factor when compared with a flat model with zero \lbd. On the other hand for the same \omeg\ and same \ob\ increasing the \lbd\ component increments the growth of perturbations. 
There is also a dependence on the geometry of the universe via the angle-distance relation which affects the value of $\Delta T/T$ observed.
This normalization is not ideal since it depends on the processing of the power spectrum of fluctuations and on the relationship between the structure we observe and the underlying density field.

\begin{itemize}
\item
The COBE-normalization 
\end{itemize}
Since the detections of CMB anisotropy other normalizations have been adopted.
On scales \simgt few degrees, CMB observations probe scales of 1000's of Mpc, inaccessible to other means of measuring large-scale structure such as redshift surveys or galaxy counts. At these large angles, the structures form part of an intrinsic spectrum of fluctuations generated through topological defects or inflation. In this linear growth regime, observations of the scalar CMB fluctuations provide a clean measure of the normalisation of the intrinsic fluctuation power spectrum.  This normalisation has been established by a number of independent CMB observations \cite{smoot,Ganga93,nature94}. Since the COBE detection \cite{smoot} it became common procedure to consider the models COBE-normalised.
This normalization is usually given in terms of the quadrupole term, $C_{2}$ or equivalently quoted in terms of the \qrms\ where:
\be
\qrms=T \sqrt{\frac{5C_{2}}{4 \pi}}, \label{qrmsc2}
\ee
where $T=2.736^{\circ}$K. 
This quantity is obtained by fitting a given model to the COBE data, according
to several techniques \cite{Gorsky94,Gorsky95,tegmarkbunn95,bond95,bunsugi94,bennett, wright94},  etc.
The relationship between this and the previous normalization is given by:
\be
C_{2} \simeq \frac{\pi}{3} \left( \frac{\ho R}{c} \right)^{4} \left( \frac{\delta M}{M} \right)^{2}_{R},
\ee
(Coles, 1995 \cite{coles95}), where $\delta M/M=\sigma_{M}$ is the mass fluctuation observed at the present epoch on a scale $R$, and is $\sigma_{8}$ when $R=8 h^{-1}$Mpc .

\subsubsection{The shape of the power spectrum} 

\begin{figure}
\begin{center}
\hspace{-4.5pt}
\psfig{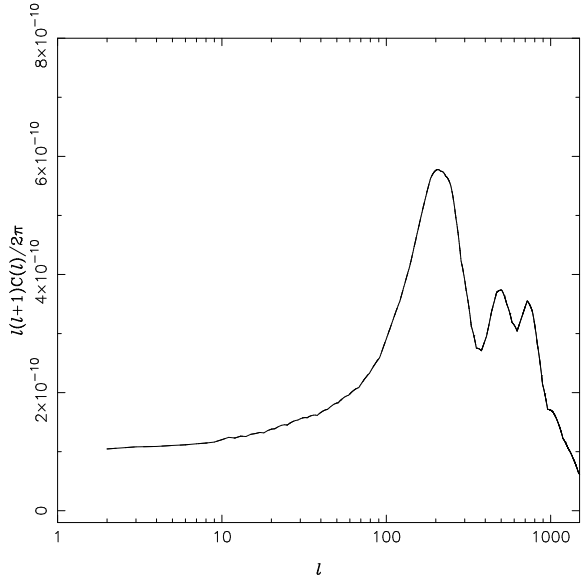}
\caption{The angular power spectrum for a standard CDM model with $\Omega=1$, $\Omega_b=0.05$, $\ho=50 \kmspmpc$ and $n=1$. The high $l$ region of the power spectrum exhibits the sequence of peaks generated by acoustic oscillations of the photon-baryon fluid at recombination while the plateau in the low $l$ region is mainly due to the Sachs-Wolfe effect. At higher $l$ we observe a damping of the fluctuations due to imperfect coupling of photons and baryons.}
\label{fig:cdm}
\end{center}
\end{figure}

\begin{itemize}
\item
Doppler peaks
\end{itemize} 
Contemporary cosmological models with adiabatic fluctuations predict a sequence of peaks on the power spectrum which are generated by acoustic oscillations of the photon-baryon fluid at recombination.
Photon pressure resists compression of the fluid by gravitational infall and sets up acoustic oscillations. The fluctuations as a function of $k$ go as 
$\cos(k c_{s} \eta_{*})$ at last scattering, where $k$ is the wavenumber of the fluctuation, $c_{s}$ is the sound speed and $\eta_{*}$ is the conformal time at recombination. This will produce a harmonic series of temperature fluctuation peaks, as shown in Fig.~\ref{fig:cdm}, 
with the $m$th peak corresponding to $k_{m}=m \pi /c_{s}\eta_{*}$.
The critical scale is essentially the sound horizon $c_{s} \eta_{*}$ at last scattering \cite{sugi95,hu,2husugi95}.
Of particular interest is the height and position of the main acoustic peak --- the so called Doppler peak. The height depends on quantities like the baryonic content of the universe, $\Omega_{b}$, and Hubble constant, $\ho$, whilst the position depends on the total density of the universe, $\Omega_{0}$ and is expected to occur on an angular scale $\sim 1\dg$. The precise form of the Doppler peak depends on the nature of the dark matter, and the values of $\Omega_0$, $\Omega_b$ and $\ho$. The scale $l_p$ of the main peak reflects the size of the horizon at last scattering of the CMB photons and thus depends almost entirely \cite{hu,kamio} on the total density of the universe according to $l_p \propto 1/\sqrt{\Omega}$. 
This relation comes from the conversion of a spatial fluctuation on a distant surface to an anisotropy on the sky. This conversion involves the spectrum of spatial fluctuations, the distance to the surface of their generation and the curvature (or lensing in light propagation to the observer). The most important effect is due to the background curvature of the universe, e.g. for an open universe a given scale subtends a smaller angle on the sky than in the flat universe, due to the fact that photons curve in their geodesics. In a $\Lambda$CDM universe (CDM model with \lbd $\neq 0$) this main peak is located at around $l \sim 220$, approximately at the same scale as in a flat CDM model, with a small dependence on $\Omega_{0}$ and $h$ \cite{pedro}. 
In conventional inflationary theory \cite{guth}, one expects the universe to be flat with $\Omega=1.0$, which can be achieved if the total mass density is equivalent to the critical density or if there is a contribution from a cosmological constant $\Lambda$ in the right amount. The height of the peak provides additional cosmological information since it is directly proportional to the fractional mass in baryons $\Omega_b$ and also varies according to the expansion rate of the universe as specified by the Hubble constant $\mbox{H}_0$; in general \cite{hu} for baryon fractions $\Omega_b \simlt 0.05$, increasing \ho reduces the peak height whilst the converse is true at higher baryon densities. In the case of topological defect models, there is also a sequence of peaks on the power spectrum for the texture model, while for the cosmic strings model only the main Doppler peak is present with no existence of the secondary peaks \cite{strings}, 
although the situation is not entirely clear yet.

\begin{itemize}
\item
Damped tail 
\end{itemize}
At higher $l$ a damping of the fluctuations occurs due to imperfect coupling of photons and baryons. The photons possess a mean free path in the baryons, $\lambda_{c}$, due to Compton scattering. Assuming a standard recombination history, the damping length approximately scales as $\lambda_{D}(\eta_{*}) \propto \eta^{\frac{1}{2}} (\Omega_{b}h^{2})^{- \frac{1}{4}}$.

\begin{itemize}
\item
Sachs-Wolfe effect
\end{itemize} 
The low $l$ region of the power spectrum is mainly contributed by the potential fluctuations in the last scattering surface, the so called Sachs-Wolfe effect \cite{sachswolfe}. This efect corresponds to red shifting of the photon as it climbs out of the potential well on the surface of the last scattering, but also to a time dilation effect which makes as to see them at a different time to unperturbed photons \cite{coles95,cdm2}. Its expression is given by:
\be
\left(\frac{\Delta T}{T}\right)_{sw}={\textstyle{\frac{1}{3}}}  \frac{\delta \phi}{c^{2}}
\ee
where $\delta \phi$ is the perturbations to the graviational potential and $c$ is the speed of light. For flat ($\Omega=1$) scale invariant ($n=1$) models with no contribution from a cosmological constant this effect gives rise to a flat power spectrum, i.e., $l(l+1)C_{l}=$constant, where the quantity $l(l+1)C_{l}$ is the power of the fluctuations per logarithmic interval in $l$. Its shape is altered if one assumes a tilted initial power spectrum with a power law, $P(k)=Ak^{n}$ or if one incorporates a cosmological constant, or assumes other than flat models. In the last two cases another source of anisotropy appears, the so-called integrated Sachs-Wolfe effect and is due to the fact that potential fluctuations are no longer time independent. An ilustration of the CMB anisotropy spectrum is given in Fig.~\ref{fig:cdm}.


\subsubsection{Calculation of the \cl's}

Theoretical calculation of the CMB anisotropies are based on linear theory of cosmological perturbations. 
Most of these calculations solve the Boltzmann equation using Legendre expansion of the photon distribution function. This method expands each Fourier mode of temperature anisotropy in Legendre series up to some $l_{max}$. This system of equations is then numerically evolved in time from the radiation dominated epoch until today. These theoretical calculations were first developed by Lifshitz (1946) \cite{lifschitz}, while Peebles and Yu (1970) \cite{peeblesyu} applied these calculations to the CMB anisotropies, considering only photons and baryons. Later other authors included e.g. dark matter (Bond \& Efstathiou 1984 \cite{be84}, 1987 \cite{be87}, Vittorio \& Silk 1984 \cite{vitsilk84}), curvature (Wilson\& Silk 1981 \cite{wilsilk81}, Sugiyama \& Gouda 1992 \cite{sugigouda92}, White\& Scott 1995 \cite{whitescot95}), tensor modes or gravitational waves (Crittenden \ea 1993 \cite{critenden93}) and massive neutrinos (Bond \& Szalay 1983 \cite{Bond93}, Ma \& Bertschinger 1995 \cite{mabert95}, Dodelson, Gates \& Stebbins 1995 \cite{dodsteb95}).
Other calculations have been done e.g. by Holtzman 1989 \cite{Holtzmann89}, Sugiyama 1995 \cite{sugi95}, Wayne \& Sugiyama 1995 \cite {hu} which used an analytic approach, Stompor 1994 \cite{stompor94}, Seljac \& Zaldarriaga 1996 \cite{uros} which method is based on integration over sources along the photon past light cone). 
The multipole coefficients $C_{l}$'s may be obtained, for example, using the solutions to the equations describing temperature fluctuations evolution by \cite{be87,cdm2}: $C_{l}=\frac{V_{x}}{8 \pi}\int_{0}^{\infty}k^{2}dk|\Delta_{Tl}(k,\tau_{0})|^{2}$,
where $\Delta_{Tl}(k,\tau_{0})$ are the coefficients of the Legendre polynomials expansion of, $\Delta_{T}(k,\tau_{0})$, the radiation intensity fluctuations, $\tau$ is the conformal time $\int dt/a(t)$ where $a(t)$ is the cosmological factor, and $k$ is the wavenumber of the Fourier mode. In particular in the case of large angular scales for $\Omega=1$ and an initial power-law form of the density fluctuations  power spectrum $P(k) \propto k^{n}$, an expression for the $C_{l}(n)$ may be used: 

\begin{equation}
C_{l}=C_{2}\frac{\Gamma(l+(n-1)/2)\Gamma((9-n)/2)}{\Gamma(l+(5-n)/2)\Gamma((3+n)/2)}  \label{eqonze}
\end{equation}

with $n<3$, $l\geq 2$, $l<40$ \cite{be87,cdm2,Scaramela88}.
 This expression is only adequate for experiments probing angular scales larger than the horizon size at recombination ($\geq 1^{\circ}$) where only Sachs-Wolfe or isocurvature effects are dominant.
In order to get the \cl's for smaller angular scales one can use of an alternative method; this consists of using the expression of the expected temperature autocorrelation function and the orthogonality property of the Legendre polynomials, which generates \cl\ for all angular scales \cite{thesis}.

In the last couple of years various groups have been producing their own CMB codes to generate the \cl's according to different approaches (G. Efstathiou, N. Sugiyama, U. Seljac and M. Zaldarriaga, M. White etc...) which are now publicly available.


\subsection{Likelihood analysis}

A most commonly used statistical technique to analyse the data is the Bayesian `likelihood function' analysis \cite{berger,nature94,lasenby94}, which we will describe in this section.
This analysis permits one to use the correlation information existing in the data and to deconvolve the switching correlations allowing one to test directly a given sky temperature fluctuation model. In order to apply this analysis one assumes a given structure formation model and considers its predicted power spectrum which defines the intrinsic angular correlation function:  

 \be
 C_{int}(\theta)=\left\langle \frac{\Delta T}{T}(\vec{n_{1}})\frac{\Delta T}{T}(\vec{n_{2}}) \right\rangle, 
 \ee
where $\vec{n_{1}}$, $\vec{n_{2}}$ are two directions on the sky such that: 
$\vec{n_{1}}$.$\vec{n_{2}}$=cos($\theta$). 

This relates with the autocorrelation function of the observations by: 

\be
C_{obs}(\theta)=\left\langle \frac{\Delta T^{obs}}{T}(\vec{n_{1}})\frac{\Delta T^{obs}}{T}(\vec{n_{2}}) \right\rangle, \label{obs}
\ee

which is equal to $C_{int}$($\theta$)$\star$$C_{beam}$($\theta$) where
$\star$ refers to convolution and $C_{beam}$($\theta$) is the autocorrelation function of the beam of the telescope. 
(see also Section~\ref{3}).  
The ACF measured between  two points on the sky $i$ and $j$ with coordinates $(\alpha,\delta)$ and $(\alpha',\delta')$ (in RA and dec), $M_{ij}$, is defined by: 

\be
M_{ij}=\left\langle \left(\frac{\Delta T}{T}\right)^{obs}_{i} \left(\frac{\Delta T}{T}\right)^{obs}_{j} \right\rangle. 
\ee

These are components of a matrix ($M_{ij}$) called the covariance matrix which in particular contains information about the beam switching. 
Adding to ($M_{ij}$) the covariance matrix for the errors ($E_{ij}$) one obtains the total covariance matrix $V_{ij}$=$M_{ij}$+$E_{ij}$. 
For Gaussian-distributed fluctuations the likelihood function for the total covariance matrix $V_{ij}$, is 

\be
 L(\Delta\ T| C_{int}(\theta))= \frac{1}{(2\pi)^{\frac{N}{2}} \sqrt{|V|}}\exp(-\frac{1}{2} \Delta\ T^{T}V^{-1} \Delta\ T).
\ee

which expresses the probability of obtaining the data set \(\Delta T=(\Delta T_{1},\ldots, \Delta T_{n})\) given some intrinsic autocorrelation function $C_{int}$.

The analysis proceeds calculating different values of \(C_{int}(\theta)\) by varying the parameters of the model and evaluating the respective likelihood. 
Searching for the maximum value of the likelihood function the most probable parameters of the model are obtained. Considering e.g. the variation of the amplitude parameter of the sky fluctuation model, two things can happen: the point of maximum can be different or equal to zero. In the first case one has a detection of fluctuations, in the second a monotonically decreasing function is obtained and there is not a detection of fluctuations in the data. The next step consists in evaluating the likelihood ratio \(L_{max}/L_{0}\) where $L_{0}$ is the likelihood evaluated at zero. This ratio gives an idea of how representative the maximum value of the likelihood function is.

The upper and lower limits on the amplitude $\sqrt{C(0)}$ of the fluctuations are obtained invoking Bayes theorem: 
\be
P(\sqrt{C(0)}| \Delta T) \propto L(\Delta\ T| \sqrt{C(0)})P( \sqrt{C(0)}),
\ee
where \(P(\sqrt{C(0)} | \Delta T)\) is the posterior probability (of obtaining the intrinsic sky fluctuations given the observed data set); $L(\Delta T| \sqrt{C(0)})$ is the likelihood (of obtaining the data set given the intrinsic sky fluctuations) and $P(\sqrt{C(0)})$ is the prior probability.
Once the form to give to the prior probability is known we are able to know the probability of obtaining the amplitude of the intrinsic sky fluctuations given the observed data set. Some Bayesians use the Jeffreys prior, uniform in log($A$) space i.e. $Pr(A)=1/A$. 
This is based on the idea that when there is no prior information available, invariance arguments imply that the prior probability density of a scale factor parameter, $A$, (which $C_{0}$ is) should be uniform in $1/A$. This poses a problem because when $A \rightarrow 0$ the $Pr(A) \rightarrow \infty$ and the posterior probability cannot be normalized. The solution could consist in considering a cutoff at low $A$. Instead we have just assumed a uniform prior such that: P($\sqrt{C(0)}$)=1 for \(\sqrt{C(0)} \geq 0\) and P($\sqrt{C(0)}$)=0 for \(\sqrt{C(0)}<0\). In this way, the area under the likelihood curve between 0 and $\sqrt{C(0)}$ gives the probability of obtaining $\sqrt{C(0)}$. 
So looking for the value of $\sqrt{C(0)}$ such that this area is 95\% of the total area one obtains a 95\% upper limit. 
In a similar way one calculates the 95\% lower limit to the intrinsic sky fluctuations (this is the equal tail prescription, ET). In a similar way, the 68\% upper and lower limits to the detected fluctuation amplitude are obtained by integrating down from the peak of the likelihood curve at plateaus of equal likelihood, until the area under the curve delimited by the upper and lower values is equal to 68 \% of the total area under the likelihood curve (this is the highest posterior density prescription, HPD.) The HPD prescription, for calculation of confidence intervals, results in somewhat smaller upper error bars and somewhat larger lower error bars then does the ET prescription.
Work has been done in order to develop methods to account for uncertainties, such as those in the beamwidth and calibration, in likelihood analysis of CMB anisotropy data \cite{GRGS}.
The beamwidth-uncertainty correction has a bigger effect for those data sets that probe a region of $l$-space where the model $C_{l}$ spectra changes rapidly with $l$. This is due to the fact that the contributions from the $\pm 1 \sigma$ beamwidth analyses no longer approximately compensate for each other.
The calibration-uncertainty correction broadens and skews the probability density distribution functions toward higher values of the amplitude ($Q_{rms-ps}$), and the calibration-uncertainty corrected likelihood has an amplitude dependent width \cite{GRGS}.


\subsection{The Gaussian auto-correlation function \label{gacf}}

At an earlier point in CMB work, the Gaussian auto-correlation function (GACF) was the tool most frequently used to describe an hypothetical sky model, and originally was intended to be used for small scales \cite{watson92,white}.
This representation is not physically realistic but is useful for comparison of the results from different experiments.
In the GACF model we assume that the intrinsic autocorrelation function is a Gaussian of amplitude $C_{0}$ and dispersion $\theta_{c}$: $C(\theta)=C_{0} \exp \left(-\frac{\theta^{2}}{2 \theta_{c}^{2}} \right)$,
even for more general autocorrelation functions is useful to define, $\theta_{c}$, by $\theta_{c}= \left(-\frac{C(0)}{C''(0)} \right)^{1/2}$.
$\sqrt{C_{0}}$ then gives an indication of the $rms$ amplitude of the intrinsic fluctuation on some coherence scale defined by $\theta_{c}$. In the angular power spectrum language, this corresponds to considering $C_{l}$ to be expressed by \cite{white}:
\be
C_{l}=2 \pi C_{0} \theta_{c}^{2} \exp( -\frac{1}{2} l(l+1) \theta_{c}^{2})
\ee
The form of the autocorrelation function after convolution with the beam from a single instrument horn, assumed to be a Gaussian, is given by the convolution of the intrinsic autocorrelation function, $C_{int}$, with the beam autocorrelation function, $C_{beam}$: $C(\theta, \sigma)=C_{int}(\theta) \star C_{beam}(\theta)$ which becomes:
\be
C( \theta, \sigma)=\frac{C_{0} \theta_{c}^{2}}{2 \sigma^{2} + \theta_{c}^{2}}
\exp \left(\frac{-\theta^{2}}{2(2 \sigma^{2} + \theta_{c}^{2}} \right). \label{eq:gausc}
\ee

Experiments observing the same intrinsic sky fluctuation do not, in general, measure the same rms signal level because this signal depends on the observational strategy. This rms signal can only be interpreted if accompanied by information about the window function of the observation. In a GACF analysis we assume a power spectrum with a Gaussian shape and take into account the peak and area of the window function. This model is a good approximation for an experiment where the power spectrum is smooth and the window function can be considered to be a Gaussian, because the two convolved spectra are very similar, in which case the quoted $C_{0}$ gives a measure of the intrinsic sky fluctuation. The sensitivity of an experiment to fluctuations on various scales is commonly described by a plot of $C_{0}$ vs $\theta_{c}$, assuming that the sky correlation function is a GACF. The experiment is most sensitive to a GACF which peaks at the maximum of its window function,and the amplitude of fluctuations to which the experiment is sensitive is parameterized by the minimum value of $C_{0}$.

In many instances experimenters now report results in terms of flat bandpower estimates, where the form of the angular power spectrum $C_{l}$ is represented by a flat spectrum over the width of a given experimental window. This is described in detail in Section~\ref{comp1}.


\section{The experiments}
\label{exper}

This Section gives an account of the current experimental status and of the main sources of contamination of data from CMB anisotropy experiments.
Usually the experiments are classified according to the angular scales they probe. The usual classification considers large angular scales as being angles larger than the horizon size at recombination i.e. $\theta \simgt 2^{\circ}$; medium scales comprehends the range $10' \simlt \theta \simlt 2^{\circ}$ and covers the region of the Doppler peaks; small scales lie in the range $1' \simlt \theta \simlt 10'$ and very small scales for $\theta \simlt 1'$.
When analysing the data from CMB anisotropy experiments one must take into account contaminants like free-free and synchroton radiation from the Galaxy which are dominant at low frequencies ($\simeq$ 10-50 GHz) and dust at higher frequencies ($\geq$ 100 GHz) \cite{laserice}. These are plotted in Fig~\ref{cont}. Another source of contamination is the possibility of discrete source contamination \cite{laserice}.
The main parameters of several recent experiments are displayed in Table~\ref{exp}.
Some of the text of this section is based upon previously published reviews by Lasenby and Hancock (see e.g \cite{lasenbyandhancock,joneslasenby}).

\subsection{Foreground emission}

\subsubsection{Diffuse Galactic emission}

\begin{figure}
\begin{center}
\hspace{15pt}
\psfig{file=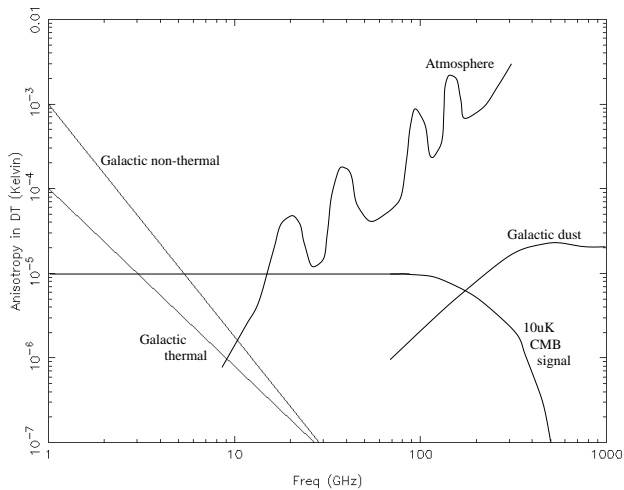,width=3.5in,clip=}
\caption{Estimates of the amplitudes of the anisotropic component of 
foreground emission (Anthony Lasenby private communication).
As a reference, a typical 10 \uk\ CMB signal is shown.} 
\label{cont}
\end{center}
\end{figure}

A detailed study of the diffuse Galactic emission has been done at low frequencies ($38 \simlt \nu \simlt 1420$ MHz), using radio telescopes \cite{lawson} and at the higher frequencies of $60 \mu $m and $100 \mu m$ using the IRAS satellite. The FIRAS and DMR instruments on board of the COBE satellite provided further information on the Galactic emission on $7\dg$ scales. In the microwave regime the emission from our Galaxy is composed of synchroton emission from cosmic ray electrons, free-free (bremsstrahlung) radiation from thermal electrons and thermal emission from dust. Fig.~\ref{cont} shows estimates (Lasenby priv. comm.) of the relative antenna temperature contributions expected from each of these diffuse Galactic foregrounds, as a function of observing frequency $\nu$.
It represents the anisotropic component of the emission at high Galactic latitude ($|b|\geq 40\dg$).
The CMB radiation has a Planckian spectral form:
\be
\Delta T_{A}=f(\nu) \Delta T=\frac{\Delta T x^{2} e^{x}}{(e^{x}-1)^{2}},
\ee
where $x=h\nu / kT$ and $\Delta T_{A}$ is the antenna temperature contribution from a blackbody temperature $\Delta T$.
The spectra of the foreground emission can be given by the relation $T_{A}=\nu^{-\beta}$, where $\beta$ is the temperature spectral index. The different frequency spectral signatures of these components allow their discrimination via observations at several frequency ranges.
The synchroton radiation is the dominant Galactic component at $\nu \simlt 30$ GHz, which results from relativistic electrons spiralling in the magnetic field of the Galaxy.
Assuming that the number density of the electrons is of the form $N(E) \propto E^{-\gamma}$, $\beta=(\gamma +3)/2$, the intensity of synchroton radiation, $I(\nu)$, is given by:  
\be
I(\nu) \propto B^{\beta-1} \nu^{2-\beta},
\ee
where B is the magnetic flux density, and $\beta$ is the temperature spectral index. The spectrum steepens from $\beta \simeq 2.75$ at 1 GHz to $\beta \simeq 2.9 $ at 15 GHz, due to increased loss of energy by electrons at higher energies. A spatial variation of $\beta$ is expected as a consequence of the spatial variation of the magnetic field \cite{banday}. So the modeling of this foreground involves the steepening of the spectrum with frequency and the spatial dependence of the frequency spectral index $\beta$.
The low frequency surveys can be used to extrapolate (see also Banday and Wolfendale 1991 \cite{banday91}) the synchroton component to other frequencies. There are low frequency surveys conducted at 408 MHz \cite{haslam}, 820 MHz \cite{berk}, and 1420 MHz \cite{reich}.
The 408 MHz is an all-sky survey while the 1420 MHz covers only the declination range $-19\dg < \delta < 90\dg$.Given the frequency range probed by these surveys it is difficult to estimate the spectral steepening at higher frequencies. Bennet \ea \cite{benet} made an attempt to account for the steepening by assuming that the spectrum of the local electrons reflects that of the electrons producing the observed synchroton emission.

The free-free emission is also a problem at $\nu < 50$ GHz and possibly higher as well. Free-free emission (bremsstrahlung) results from the acceleration of electrons (when they interact with the warm ionized component of the Galaxy) in the electric field of the ions. Its spectral index is a weak function of frequency \cite{benet} and is given by:
\be
\beta_{ff}=2+ \frac{1}{10.48+1.5 \ln (T_{e}/8000{\rm K})- \ln \nu ({\rm GHz})},
\ee
where $T_{e}$ is the electron temperature and $\beta \simeq 2.1$ at the Tenerife experiment frequencies. This component of emission is large within the Galactic plane, but at $\nu \simlt 15$ GHz and at high Galactic latitude it is expected to be much lower than the synchroton emission. It is thought that the free-free emission reaches $\simeq$ the same level of the synchroton at frequencies $\simeq$ 20-30 GHz, and to be larger for higher frequencies, although the accurate amplitude of this Galactic foreground is not yet well understood.

At higher frequencies, larger than $\simeq$ 100 GHz, the contribution to foreground emission comes from the Galactic dust emission. The modeling of this foreground is difficult because it requires the knowledge of the temperatures, emissivities and spatial distributions of the dust grains in the Galaxy. Before the publication of COBE data, Banday and Wolfendale 1991 \cite{bandaywolf} concluded that IRAS observations provided useful information at high frequencies but its extrapolation for frequencies where the CMB is observable is difficult given the uncertainties in the dust model.
Bennet \ea 1992 \cite{benet}, using empirical fits to the COBE FIRAS data estimated the dust contribution to the COBE observing frequencies and found that the $rms$ dust signal was less than 8 \uk\ at 90 GHz.
Dust emission increases with increasing frequency with a spectral index $\beta \simeq -1.5$.

\begin{itemize}
\item Discrete source emission
\end{itemize}

Another source of contamination of CMB observations is the discrete source emission. Most of these sources have a flat or falling spectra such that their flux decreases with increasing frequency. The identification of the brightest sources at low frequencies is made using the existing catalogues, but in some cases it is necessary to use separate high resolution surveys.
The antenna temperature contribution for a source of flux, S, in a single beam is given by:
\be
T_{A}=\frac{A_{eff} S}{2k},
\ee
where $A_{eff}=\pi \left( \frac{c}{2 \nu \theta_{\rm FWHM}} \right)^{2}$,
is the effective area of a single beam. Even for a flat spectrum i.e. $S \propto \nu^{0}$, $T_{A}$ decreases with frequency as $\nu^{2}$, and is inversely proportional to the square of the beamwidth. On scales $\theta < 0.5\dg$ at $\nu <20$ GHz, the foreground radio sources are important contaminants of CMB observations. Franceschini \ea \cite{franceschini89} concluded that for $\theta < 1\dg$ and 20 GHz$\leq \nu \leq$200 GHz the contribution to $\Delta T / T$ is just below $10^{-5}$. The Kuhr \ea catalogue \cite{kuhr} of radio sources provides information about surveys ranging in frequency from 26 to 90 GHz.

\begin{itemize}
\item Atmospheric emission
\end{itemize}

Atmospheric emission causes serious problems to CMB observations. Water vapour emission lines dominate at 22 and 182 GHz, while oxygen line emission is significant at 60 and 118 GHz. These emission lines define the range of frequencies of the atmospheric windows through which CMB observations can be made. Experiences with differential observing techniques are only affected by the anisotropic component of the atmospheric emission.
Fig.~\ref{cont} shows an estimate of this anisotropic component at a good ground-based observatory. The atmospheric emission increases with frequency and is a serious obstacle to CMB observations with increasing frequencies.

\subsection{Large angular scales}

The large scale CMB anisotropy is mainly due to the so-called Sachs-Wolfe effect (see Section~\ref{basic}). Fluctuations in scales larger then the horizon size at recombination retain their primordial characteristics since they have not been changed by any causal processes inside the horizon before recombination. So the CMB power spectrum mirrors that of the initial seed perturbations and therefore reflects the primordial unprocessed power spectrum.  
Observations at these scales allow us to determine the normalization and slope of the primordial power spectrum. They may also be used to distinguish between the adiabatic and isocurvature fluctuations. 
These tasks can be complicated by the existence of a gravity wave background. 
At these scales it is not expected to obtain information about the Gaussianity of the fluctuations since the beam will average over the number of defects and the central limit theorem states that the result will be Gaussian.
There are three main experiments operating on these scales: COBE, Tenerife, MIT/FIRS.

\small
\begin{table}
\begin{center}
\begin{tabular}{|c|c|c|c|} \hline         
Experiment & Institution & $\theta_{beam}$/  & $\nu$ (GHz) \\
 & & $\theta_{throw}$($\dg$) & \\\hline
DMR-COBE (S) & NASA & 7 & 31, 53, 90  \\
Tenerife (G) & NRAL/MRAO/IAC & 5/8 & 10, 15, 33 \\
MIT/FIRS (B) & GSFC/Chicago/\ldots & 3.8 & 180 +3 higher \\\
ACME/HEMT (G) & UCSB & 1.5/2.1 & 30/40 \\
MAX (B) & Berkeley/UCSB & 0.5/1.0 & 110, 180, 270, 360 \\
MSAM (B) & GSFC/Chicago/\ldots & 0.5/0.6 & 180 + 3 higher \\
White dish (G) & CARA & 0.18/0.47 & 90 \\
Python (G) & CARA  & 0.75/2.75 & 90 \\ 
Saskatoon (G) & Princeton & 1.5/2.45 & $26-36$ \\
ARGO (B) &  Rome/Berkeley & 0.9/1.8 & 150 + 3 higher \\
CAT (I) & Cambridge & 0.25 & 13.5, 15.5, 16.5 \\
\hline
\end{tabular}
\end{center}
\caption{Some recent CMB anisotropy measurements. (S) Satellite, (G) Ground, (B) Balloon, (I) Interferometer, ($\theta_{beam}$) Beamwidth, ($\theta_{throw}$) Beamthrow.}
\label{exp}
\end{table}

\subsection{Medium angular scale observations}

At these angular scales models predict larger amplitudes of the CMB fluctuations arising from the Doppler peaks (see Section~\ref{basic}). These oscillatory features are very sensitive to the details and are model dependent, therefore their observation should allow good constraints to be established on model parameters. Comparing these experiments to those on larger angular scales one may separate the scalar fluctuations from any possible gravity wave background. At these scales the effect of the cosmic variance is small but a new source of uncertainty arises due to the small size of the sky areas probed by these experiments, the sample variance. According to Coulson \ea \cite{coulson}, at these scales, it should be possible to detect non-Gaussianity induced by defect models. Reionization of the Universe at some epoch can smear out these fluctuations to scales depending on the epoch of reionization. If it happened early enough it could erase perturbations up to scales of less than few degrees.
There are several experiments operating on these scales SP/ACME-HEMT, Saskatoon, ARGO, Python, MAX, MSAM, etc.

\subsection{Small and very small angular scales}

At these angular scales the amplitudes of the fluctuations are expected to be reduced due to damping effects such as the Silk damping and damping due to the finite thickness of the last scattering surface (see Section~2).
They may be further reduced by secondary scattering processes and damping mechanisms, although if there has been reionization the generation of CMB fluctuations at small scales is expected (Ostriker-Vishniac effect).
These experiments are sensitive to imprints on the CMB from the seeds of galaxies and clusters of galaxies. Observations at these scales give information about the non-linear physics of galaxy formation and about the thermal history of the universe.
They may be compared with other experiments in order to constrain the angular power spectrum of a given model.
Examples of experiments operating on these scales are, among others, OVRO, VLA, etc.

\subsection{Future Experiments}

In the near future it is expected that new instruments will be built with better accuracy and a broader range of frequency coverage, providing improved quality CMB data.
For balloon experiments one can improve the quality of the data through the use of long-duration balloon flights, e.g. launched in Antartica, and circling the Pole, and pixels sampled, and  the use of arrays of detectors in order to extend the frequency range observed.
Examples of these are Boomerang, Maxima, and TOPHAT.
On the ground a great improvement is expected from interferometers, one example is the Very Small Array (VSA), to be built by Cambridge and Jodrell Bank in the U.K., and to be sited in Tenerife, which is expected to be operational by the year 2000.
The CAT is a prototype for this more advanced instrument.
The VSA is expected to give detailed maps of the CMB anisotropy with a sensitivity $\simeq$ 5 \uk\ and comprehending a range of angular scales from $10'$ to $2\dg$, and covering a frequency range of 28 and 38 GHz.
This instrument uses an interchangeable T-shaped configuration of 10-15 horn elements and simulations have shown that such a configuration is able of attaining a good sensitivity over the range of angular scales planned to be used.
It is also expected to measure the values of $\Omega$ and \ho with an accuracy of better than 10\% due to the good accuracy expected over the region of the first and secondary Doppler peaks. Simulations have shown that this instrument will be quite sensitive to non-Gaussian features on these angular scales. 
For future satellite experiments one has the Microwave Anisotropy Probe (MAP) and Planck Surveyor satellite.
MAP has been selected by NASA as a Midex mission and is expected to be launched between 1999 and 2001 while Planck has been selected by ESA as an M3 mission and will be launched probably around 2005.
A very important characteristic of satellite experiments is that they are not affected by the problems caused by the atmosphere. Consequently a satellite is capable of full-sky coverage and has the potential to map features on large angular scales ($> \sim 10\dg$). On the other hand it has more problems in reaching resolution on smaller angular scales due to the limitations imposed on the
 dish size. 
The MAP median resolution of its channels is around 30 arcmin while the best angular resolution is 18 arcmin, in its frequency channel at 90 GHz.
Consequently it may have problems in determining the shape of the first and almost certainly of the secondary Doppler peaks of the angular power spectrum.
Planck is expected to attain a resolution around 4 arcmin, with a median resolution of its 6 channels of about 10 arcmin.
Consequently it will be possible to determine the angular power spectrum with good accuracy including the secondary peaks, and therefore the determination of the cosmological parameters with good accuracy.
With the good angular resolution of Planck surveyor it is expected to obtain a joint determination of $\Omega$ and \ho\ to 1\% accuracy.
All these, balloon, ground-based and satellites experiments to come constitute a good improvement of CMB data and represents an important step towards understanding the characteristics of our Universe.


\section{The Tenerife experiment}
\label{ten}

We here give a summary of the analysis and interpretation of results obtained on relatively large angular scales of $\sim 5\dg$ by the Tenerife experiment.
Observations of fluctuations in the Cosmic Microwave Background (CMB) have been widely recognized to be of fundamental significance to cosmology, offering a unique insight into the physical conditions in the early Universe.
As we have seen the amplitudes and distribution of such fluctuations provide critical tests of the origin of the initial perturbations from which the structures seen today have formed. On scales \simgt few degrees, CMB observations probe scales of 1000's of Mpc, inaccessible to conventional astronomy. At these large angles, the structures form part of an intrinsic spectrum of fluctuations generated through topological defects or inflation. In this linear growth regime, observations of the scalar CMB fluctuations provide a clean measure of the normalisation of the intrinsic fluctuation power spectrum.  This normalisation has been established by a number of independent CMB observations \cite{smoot,Ganga93,nature94}. In many theories, tensor CMB fluctuations from a background of gravitational waves are also expected to be significant on these large scales, and measuring the slope of the power spectrum offers the potential to constrain this contribution to the CMB anisotropy \cite{nature94, steinhardt,crittenden}. A comparison of the large-scale anisotropy results with those on medium scales can also be used to separate the scalar and tensor components under the assumption of a specific cosmological model.

The Tenerife CMB experiments were initiated in 1984, with the installation of the first 10 GHz switched-beam radiometer system at the Teide Observatory on Tenerife Island. A subsequent programme of development has led to the present trio of independent instruments working at 10, 15 and 33 GHz. The ultimate objective is to obtain three fully sampled sky maps covering some $\sim 5000$ square degrees of the sky at each frequency and attaining a sensitivity of $\sim 50$ $\mu$K at 10 GHz, and $\sim 20$ $\mu$K in the two highest frequency channels. Drift scan observations have been conducted over a number of years covering the sky area between Dec=+30\degg~and +45\degg.  The deepest integrations have been conducted in the Dec=+40\degg~region and resulted in strong evidence for the presence of individual CMB features (Hancock \et 1994 \cite{nature94}).

Davies \et 1995 \cite{davies95} (hereafter Paper I) described the performance of the experiments and gave an assessment of the atmospheric and foreground contributions to our data at Dec=+40\degg; Hancock \et 1996 \cite{me96} (hereafter Paper II) analysed in detail the results and cosmological implications of such observations. 
Here we present a summary of the analysis and results published in Hancock \et 1996 \cite{me96}. 
Section~\ref{scan40} gives a brief description of the observational strategy. 
In Section~\ref{anstat} we use one of the several statistical methods to calculate the level of the detected signals and their origin. A statistical comparison with the results of the COBE DMR two-year data conducted by Hancock S. and Tegmark M. is mentioned in Section~\ref{comb} and used to place limits on the spectral index of the primordial fluctuations.

\subsection{The scans at Dec $+40\dg$ \label{scan40}}

\subsubsection{Observations}

Observations were conducted at the three frequencies 10, 15 and 33 GHz by drift scanning in right ascension at a fixed declination of 40\degg. The measurements were made independently at each frequency, using separate dual-beam radiometer systems as described in Paper I.
The three instruments are physically scaled so as to produce approximately the same beam pattern (FWHM$\sim 5\degg$) on the sky, thus allowing a direct comparison of structure between frequencies. A characteristic triple beam profile (switching angle 8\deg 1) is obtained by the combination of fast switching (63 Hz) of the horns between two independent receivers and secondary switching (0.125 Hz) provided by a wagging mirror. We make repeated observations of the sky, binning the data in 1\degg~intervals in RA and stacking them together in order to reduce the noise as compared with individual measurements. As a consequence of using two independent channels, the receiver noise contribution to the final data scans is reduced by a factor $\sqrt{2}$ compared with single channel observations.
The data considered here are those presented in the preliminary report by Hancock \ea (1994) \cite{nature94} and Hancock \ea (1996) \cite{me96}.
We restrict our analysis to the RA range $161\dg-230\dg$ corresponding to Galactic latitude $b>56\dg$. At these high latitudes foreground emission from the Galaxy is expected to be at a minimum. 
This sky region has also been selected (see Paper I) to be free from discrete radio sources above the 1.5 Jy level at 10 GHz.
The contribution of unresolved radio sources is expected to be $\Delta T/T \le 5\times10^{-6}$ at 15 GHz and significantly smaller at 33 GHz \cite{aizu87,franceschini89}.
The revised sensitivities per beam-sized area are 61, 32, 25 and 20 $\mu$K at 10, 15, 33 and 15+33 respectively. 
For details see Paper II.

\subsubsection{Reliability of the detected signals}

Determination of the amplitude of the CMB component of the structure requires one to consider the contributions of random noise and foreground signals to the observed data scans. The former has its origin in the thermal variations in the receivers and in the fluctuating component of the atmosphere, whilst the latter consists primarily of free-free and synchrotron emission in the Galaxy, plus emission from the Sun and Moon. Of these effects, only the Galactic emission remains constant from day to day at a given frequency. In Paper I it was estimated a maximum Galactic contribution of $\Delta T_{rms}=4 \mu$K in the results at 33 GHz. An improved separation between the Galactic and the cosmological signal at each frequency has been under consideration via a new maximum entropy based method to reconstruct the intrinsic sky fluctuations as developed in Jones \ea \cite{aled97}.

The presence of correlated atmospheric component was taken into consideration in paper II, via a new split of the 33 GHz data into two subsets $X$ and $Y$ such that both channels of a given scan are included in the same data subset. 
An analysis similar to that presented in Hancock \ea (1994) \cite{nature94} gives an astronomical signal ($\sigma _s^2=\sigma_{(X+Y)/2}^2-\sigma_{(X-Y)/2}^2$) with an amplitude $\sigma _s=43 \pm 12$ $\mu$K.
The value of the signal quoted in Hancock \et \cite{nature94} was $\sigma_{old}=49 \pm 10$ $\mu$K; the difference with the improved estimation is certainly due to the subtraction of the atmospheric signal. The difference between both estimates ($\sigma=(\sigma_{old}^2-\sigma _s^2)^{1/2}$) is $\sim 23$ $\mu$K which is our best assessment of the atmospheric contamination in the analysis based on the addition and difference of the 33 GHz data; this value is in good agreement with estimates obtained using other methods.
For details see Paper II.

\subsection{Statistical analysis \label{anstat}} 

The estimates of the astronomical signals made in the previous section can be improved on using a detailed statistical analysis. Here we use a likelihood analysis and take into account the contribution of the correlated atmospheric noise by enlarging the error bars on the stacked scans. 
These modified scans are also used to study the presence of common features at 15 and 33 GHz by the calculation of the cross-correlation function (for details see Paper II).
Here we will focus our attention to the likelihood analysis technique.

\subsubsection{Likelihood analysis}

\label{likelihood}

This analysis takes into account all the relevant parameters of the observations: experimental configuration, sampling, binning, etc. (see Section~\ref{basic}).  The combination of the atmospheric and instrumental noise can be modeled by a Gaussian distribution uncorrelated from point to point, (Paper II), which implies that the noise only contributes to the diagonal terms of the covariance matrix. We also assume that the astronomical signal is described by a Gaussian random field and therefore our results correspond to a superposition of Gaussian fields in which all their statistical properties are specified by the covariance matrix, which takes into account the full correlation between the data points. The likelihood analysis procedure is to vary the parameters of the model to obtain different values of $C_{int}(\theta)$ and then to calculate the likelihood of observing the data set $\Delta T$. Then the most probable model parameters are determined and a detection of fluctuations will show itself as a peak in likelihood away from zero (see Section~\ref{basic}).

\begin{itemize}
\item
Fluctuations with a Gaussian auto-correlation function (GACF)
\end{itemize}

We have analyzed our results for two hypothetical sky models, the first of which corresponds to a signal described by a Gaussian auto-correlation function (ACF) with amplitude $\sqrt{C_0}$ and width $\theta _c$ (see Section~\ref{basic}). This is not a realistic physical scenario but has been used widely in the past \cite{davies87,readhead89,watson92} because it provides for an easy comparison between the results of experiments with different configurations. 
The intrinsic ACF for these models is a Gaussian of amplitude $\sqrt{C_0}$ and dispersion $\theta _c$,
which is modified accordingly by our triple beam filtering (\cite {watson92}, see Section~\ref{basic}).  Our instrument is sensitive over a range of coherence angles $1\dg \simlt \theta_c \simlt 10\dg$, attaining peak sensitivity for a coherence angle of 4\degg.

\begin{figure}
\vbox{%
\hspace*{1in}
\psfig{file=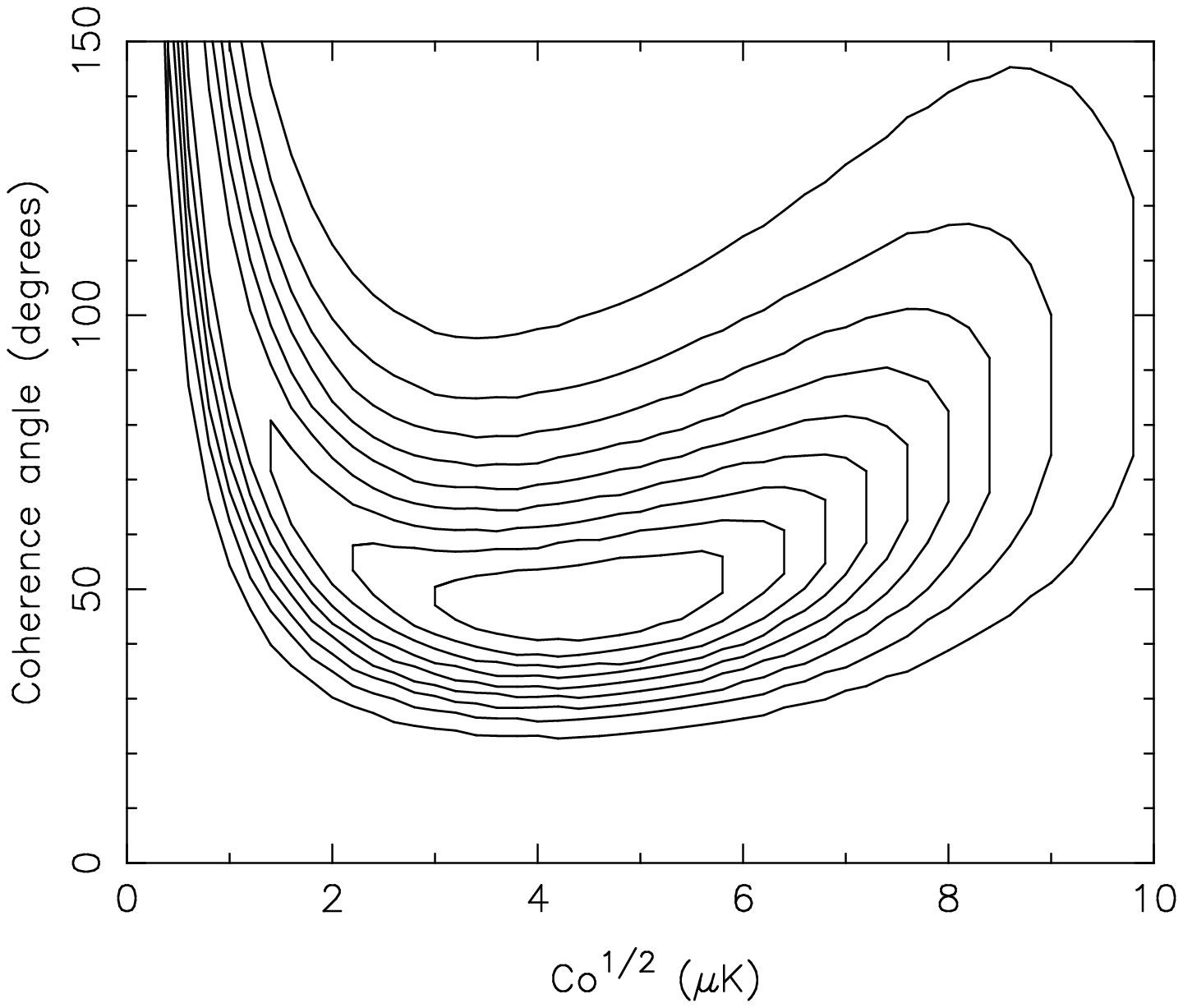,width=4.3in,angle=270,clip=}
\hspace*{0.5in}
\psfig{file=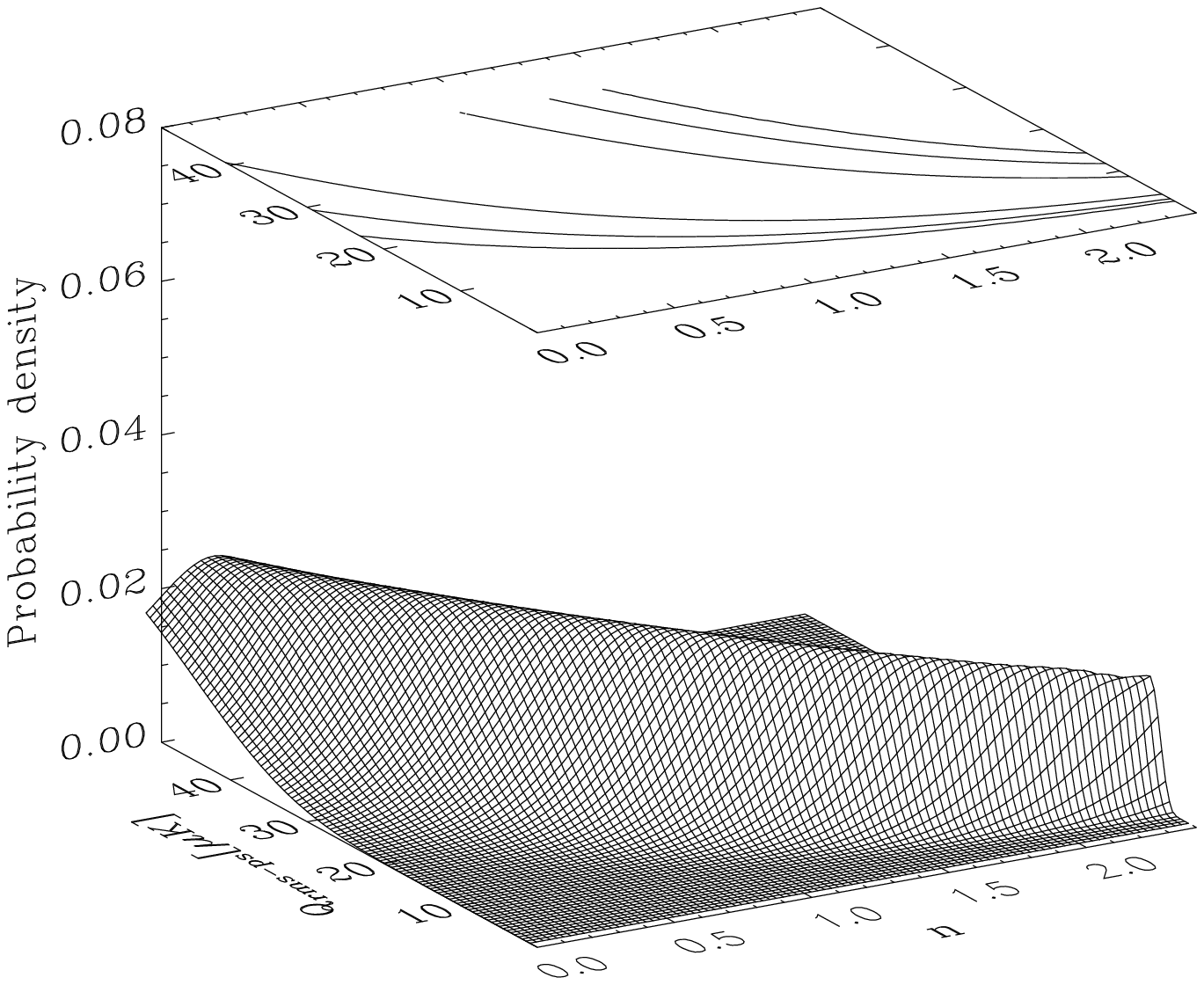,width=3.9in}} 
\caption{Top Figure: Contour levels of equal likelihood for the 15+33 scan in the case of a Gaussian shaped ACF. The contours correspond to 10, 20, 30, 90 \% of the total probability distribution. The Tenerife configuration obtains maximum sensitivity for coherence angles in the range $2\dg \simlt \theta \simlt 6\dg$; structure is clearly detected at $\sim 50 \mu$K over this angular scale range. 
Bottom Figure: The two-dimensional normalised likelihood surface as a function of the spectral index $n$ and the normalisation \qrms for the Tenerife data (in the case of fluctuations with a power law spectrum). The projected contours are at 68 \%, 95 \% and 99 \% confidence.}
\label{Figure3}
\end{figure}

We have applied this analysis to the latest data at declination $40^{\circ}$ described in Section~\ref{scan40} and Paper II. We have analyzed the $X$ and $Y$ subsets for 15 and 33 GHz, the total stacked scans at these two frequencies, and our best scan 15+33. The results for $\theta_c=4\dg$ are presented in the third column of Table~\ref{tabres}. 
The amplitude of the intrinsic signal corresponding to the maximum likelihood is given, along with the one-sigma confidence bounds calculated in a Bayesian sense with uniform prior. All results look consistent with clear detections at the two to three sigma level, and mean values of the signal slightly smaller than those presented in Hancock \et \cite{nature94}, due to the improved estimate of the error bars in the stacked scans (see Section~\ref{scan40} and Paper I). Fig.~\ref{Figure3} presents the contours of equal probability for the 15+33 scan. We see a well defined point of maximum likelihood at $\theta_c\sim 4$\degg~and $\sqrt{C_0}\sim 50$ $\mu$K.

\begin{itemize}
\item
Fluctuations with a power law spectrum
\end{itemize}

The second model considered here is more interesting from a cosmological viewpoint. It corresponds to the prediction of the power law form ($P(k)\propto k^n$) for the spectrum of the primordial fluctuations (see Section~\ref{basic}). Considering only the Sachs-Wolfe part of the spectrum of the fluctuations the intrinsic ACF can be expressed as 
$C(\theta)=\frac{1}{4 \pi}\sum _l (2l+1) C_l^{(S)} P_l(\cos \theta)$ with $C_l^{(S)}=C_2^{(S)}\frac{\Gamma[l+(n-1)/2]\;\Gamma[(9-n)/2]}{\Gamma[l+(5-n)/2]\;\Gamma [(3+n)/2]}$
where the sum is extended to the multipoles $l\lta 60$ which corresponds to the range of angular sensitivity of our experiments. For $l \gta 20$, standard models predict additional contributions to the CMB anisotropy, as one moves into the low $l$ tail of the CMB Doppler peak. Hence fitting for the Sachs-Wolfe term alone 
to CMB data on these scales can lead to the derived values for $n$ being increased by as much as 10\% over the true primordial value. This point should be borne in mind when comparing the limits on $n$ from the Sachs-Wolfe term (Section~\ref{comb}) to those from a fit to a full CDM type functional form as in Section~\ref{comp2}. For a given value of the spectral index $n$, the intrinsic ACF is a function only of \qrms.

This analysis was applied to the recent data described in Section~\ref{scan40}. Fig.~\ref{Figure3} shows the likelihood surface as a function of \qrms and $n$ for the 15+33 scan. The peak likelihood forms a ridge displaced from zero in \qrms and corresponds to a $3-4$ sigma detection of structure for each value of $n$ considered. The shape of the surface implies that all values of $n$ in this range are equally likely. This is predominantly a consequence of our observing technique which samples only a small angular range of the spectrum of fluctuations. Thus whilst our observations provide a good measure of the fluctuation power on $\sim 4\dg$ scales, they do not in themselves contain sufficient information about the distribution of power with angular scale to allow a useful determination of the spectral slope: for this one must compare with experiments on other angular scales (see Section~\ref{comb} and Sections~\ref{comp1} and \ref{comp2}). For the specific case of a \hz spectrum ($n=1$) the results of the likelihood analysis are given in column two of Table~\ref{tabres}. The normalisation \qrms corresponds to the maximum of the likelihood function and the confidence intervals are at 68 \%, calculated in the standard Bayesian manner using uniform prior. We see that in general the results are consistent and agree with a global normalization of the quadrupole $\qrms \sim 20-25$ $\mu$K. Our best estimate for \qrms of $22^{+10}_{-6} \mu$K from the 15+33 scan is reduced over the value of $26\pm 6 \mu$K previously reported due to our now having properly accounted for the correlated atmospheric noise.

\begin{table}
\begin{center}
\caption{Results of the likelihood analysis for a Harrison-Zel'dovich spectrum of fluctuations (second column) and for a Gaussian ACF (third column).}
\vskip 0.2 cm
\begin{tabular}{ccc}
& &  \\
$\nu$ (GHz) &  \qrms ($\mu$K) & $\sqrt{C_0}$ ($\mu$K) \\
\hline 
& & \\
15A & $27^{+16}_{-16}$ & $54^{+37}_{-30}$ \\
& & \\
15B & $17^{+9}_{-17}$ & $24^{+21}_{-24}$ \\
& & \\
15  & $21^{+12}_{-9}$  & $44^{+26}_{-19}$\\
& & \\
33A & $22^{+14}_{-10}$ & $45^{+32}_{-24}$ \\
& & \\
33B & $28^{+12}_{-9}$ & $57^{+28}_{-25}$ \\
& & \\
33  & $24^{+11}_{-8}$ & $49^{+27}_{-17}$ \\
& & \\
15+33 & $22^{+10}_{-6}$ & $48^{+21}_{-15}$ \\
& & \\
\end{tabular}
\label{tabres}
\end{center}
\end{table}

\subsection{Statistical comparison with COBE DMR}

\label{comb}

A comparison between the results of different CMB experiments offers the opportunity to check independent measurements, to extend the range in frequency and angular scale, to constrain cosmological models and, if the sensitivity of the experiments is sufficient, to compare features. There are several experiments operating on angular scales of a few degrees: MIT \cite{Ganga93}, COBE  \cite{smoot,bennett94}), RELIKT \cite{strukov93}, ARGO \cite{argo} and Tenerife. The first comparison between independent CMB observations was made by Ganga \et (1993) \cite{Ganga93} who found a clear correlation between the results of the first year of COBE DMR observations and those of the MIT experiment. De Bernardis \et (1994b) \cite{debernardis} have also made a statistical comparison between the amplitude of the signal reported for ARGO and that of the COBE DMR first year results, from which they constrain the spectral index to be $0.5\le n \le 1.2$ in the absence of any gravity wave background. In the preliminary report (Hancock \et 1994 \cite{nature94}) the amplitude of the signal detected on $\sim 5\dg$ scales in our Dec=+40\degg~data was compared with that on $\sim 7\dg$ scales for the first year of COBE data.
It was found that both results were consistent with inflationary models ($n\simgt 0.9$) but with a favoured spectral index of $n=1.6$ (this value decreases if we use in the comparison the new astronomical signal corrected for the atmospheric contamination). Lineweaver \et (1995) \cite{lineweaver95} have presented the first direct comparison of CMB features between the two-year COBE DMR data and the Tenerife Dec=+40\degg~observations, confirming the agreement in the level of the normalization of both experiments and providing clear evidence for the presence of common hot and cold spots in both data sets. 
Here we give a brief account of a new comparison carried out by S. Hancock and M. Tegmark \cite{me96}. 
This comparison is different to that conducted in Hancock \et (1994) \cite{nature94} in that they use  more rigorous comparison technique that utilises the likelihood function to incorporate fully the effects of cosmic and sample variance, random noise and the interdependence of the model parameters. In addition to the improvements from this revised analysis, the new results also reflect the increased sensitivity of the COBE data after two years of observing, along with the more accurate estimate of the cosmological signal in the Tenerife data.

\subsubsection{Properties of the two data sets}

The instrumental profile of the COBE data is described approximately by a Gaussian beam with FWHM$\sim 7\degg$. As can be seen in Fig.~1 of Watson \et (1992) there is a range of angular scales to which the COBE DMR and Tenerife experiments are both sensitive. Measurements taken by COBE cover the full sky, but to determine the CMB fluctuations the region of the Galactic plane has been excluded  ($|b|\le 20$\degg) thereby introducing a degree of uncertainty in estimating the properties of the global field; this effect is commonly termed sample variance \cite{samvariance} (Section~2). The uncertainties in the COBE two year results are dominated by the effect of cosmic variance {\it i.e.} the fact that our stochastic theory describes the Universe as a particular realisation of a random field (Section~2). Together the cosmic and sample uncertainties form an intrinsic limitation of the COBE experiment since unlike random errors they are not reduced by increased integration time.
The two year COBE data have been analysed independently by a number of authors (\eg \cite{banday94,bennett94,Gorsky94,wright94}). All find evidence for statistically significant structure at an amplitude consistent with that of $30\pm5$ $\mu$K \rms on a $10\dg$ scale announced by Smoot \ea (1992) for the first year data. The best fit values for $n$ and \qrms depend on the precise analysis techniques employed, but are generally consistent with the values of $n=1.10 \pm 0.29$, $\qrms=20.3\pm 4.6$ $\mu$K found by Tegmark and Bunn (1995) \cite{tegmarkbunn95} for the combined 53 and 90 GHz data with the quadrupole included.
In the case of the Tenerife experiment the double-switching scheme removes the contribution of low order multipoles decreasing the cosmic variance of the signal on these large scales; the major source of uncertainty is produced by the partial sky coverage (sample variance) and the instrumental noise. The region observed by the Tenerife experiments covers $\sim 5000$ square degrees but here we have limited our analysis to our region of deepest integration at high Galactic latitude which constitutes a sample $\sim 500$ square degrees.
For such a region the uncertainties due to the partial sky coverage dominate over the intrinsic variance by a factor $\sim 10$ \cite{samvariance,thesis} and the combined uncertainty is approximately of the order of that introduced by the instrumental noise in the 15+33 scan.

What is required is a data analysis technique that allows the joint probability of any combination of the model parameters to be calculated and which implicitly takes into account random errors and cosmic and sample uncertainties.  The Bayesian approach using the likelihood function as described in Section~\ref{anstat} attempts to do precisely this.  The likelihood function peaks at the most likely parameters (the best estimate of the true values if the likelihood function is unbiased) and has some distribution which through Bayes theorem is representative of the combined effects of the cosmic, sample and random uncertainties.  The issue of how well this distribution reflects the true uncertainties is addressed 
by comparison of the Bayesian probability distribution with that obtained from direct Monte Carlo simulations of the data (see \cite{thesis}). The Bayesian and frequentist approaches are found to be consistent for the Tenerife data and to a good approximation the likelihood function is also seen to be an unbiased estimator of the model parameters.  Consequently the likelihood surface for the joint Tenerife and COBE data set provides the definitive means of comparison of the observations under some assumed sky model.

\subsubsection{The Tenerife-COBE likelihood function}
\label{likejoint}

S. Hancock and M. Tegmark applied the likelihood analysis to the COBE two year data and the Tenerife 15+33 scan, assuming a power law model with free parameters $n$ and $\qrms$. The COBE Galaxy-cut two-year map consists of 4038 pixels, whilst the Tenerife Galaxy cut (RA $161\dg-230\dg$) scan contains 70 pixels, requiring a 4108 $\times$ 4108 covariance matrix for a joint likelihood analysis of the data. For details about the implementation of the joint Tenerife-COBE likelihood function see Hancock \ea 1996 \cite{me96}.

\begin{figure}
\vbox{%
\psfig{file=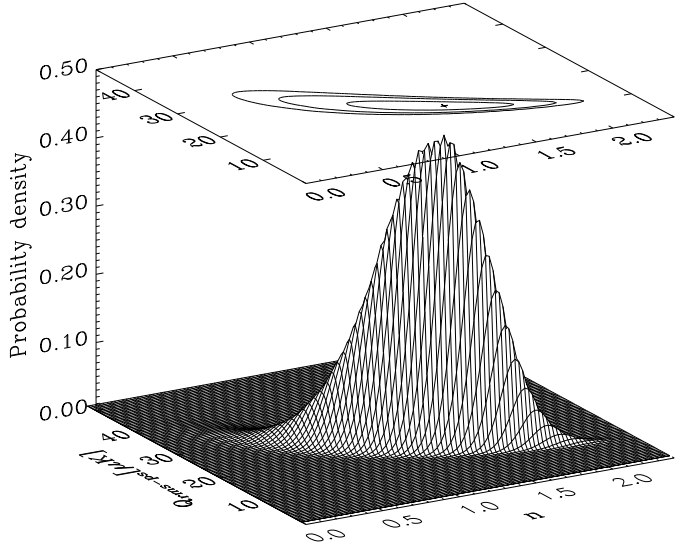,width=4.5in}
\hspace*{1in}
\psfig{file=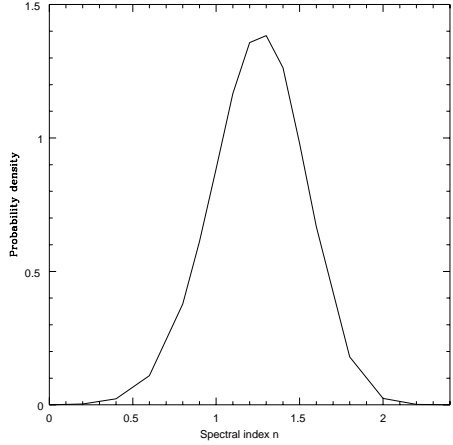,width=2.5in}}
\caption{Top Figure: Constraints on the quadrupole \qrms and on the spectral index $n$ of fluctuations obtained from a joint likelihood analysis of the Tenerife data and COBE DMR two-year data. The contour levels represent 68 \%, 95 \% and 99 \% of the region of joint probability. The peak of the distribution lies at $n=1.37$, $\qrms=16.1 \mu$K and is identified by the cross.
Bottom Figure: The marginal likelihood for the spectral index $n$ as obtained from the joint analysis of the Tenerife and COBE data. The spectral index is seen to lie in the range $1.02 \le n \le 1.62$ at 68\% confidence, with a best fit value of $n=1.33$.}
\label{Figure6}
\end{figure}

For our Tenerife data, the best estimate of the cosmological signal is obtained from the 15+33 combined scan after correction of the error bars for the correlated atmospheric noise term.  The possible contribution of Galactic signals has been estimated from the 10 GHz data to be less than 4 $\mu$K at 33 GHz and has not been considered in the current comparison. The normalised likelihood function for this scan, as plotted in Fig.~\ref{Figure6}, represents the joint probability of obtaining a given combination of $n$ and $\qrms$. On its own, the Tenerife configuration provides less leverage on the slope of the spectrum than the COBE satellite. This is because the Tenerife experiment is  insensitive to the largest angular scales, and because the one-dimensional shape of the  dec$+40^\circ$ strip makes it difficult to separate the power contributions from different scales. In other words, a narrow strip corresponds to wide window functions in $\ell$-space, with considerable aliasing of small-scale power onto larger scales (just as the case is with one-dimensional ``pencil beam" galaxy surveys). 
As a result, the Tenerife data can be equally well fit by a range of $n$ and $\qrms$ values, resulting in a likelihood ridge in $(n,\qrms)$-space with minimal discriminatory power for the parameter $n$. In contrast, the COBE observations are sensitive to the slope to the extent that the likelihood surface is peaked in the $n$-dimension. Combining the COBE information with the Tenerife data improves the situation in two ways: it extends the lever arm on the spectral slope from the COBE scales down to the $4\dg$ scale of Tenerife, and in addition eliminates the above-mentioned aliasing problem, since the joint data set is no longer one-dimensional.

Fig.~\ref{Figure6} shows the confidence contours obtained from Bayesian integration under the combined COBE and Tenerife likelihood surface assuming a uniform prior. The 68\% joint confidence region in $(n,\qrms)$-space encloses a region from 0.90 to 1.73 in $n$ for $\qrms$ in the range from $12.1$ to $22.9$ $\mu$K, with the peak at $n=1.37$, $\qrms=16.1\mu$K. Marginalizing over $\qrms$ with a uniform prior, one obtains the probability distribution for $n$ as given in Fig.~\ref{Figure6}, corresponding to $n=1.33\pm 0.30$ at $68\%$ confidence. The resulting limits on the normalization, conditioned on $n=1$ as is customary, are $\qrms=21.0\pm 1.6$. The corresponding results in Tegmark and Bunn (1995) \cite{tegmarkbunn95}, hereafter ``TB95" using just the COBE data, and including the weak correlated noise term \cite{lineweaver95}, were $n=1.10\pm 0.29$ and $\qrms=20.3\pm 1.5$, with the peak likelihood located at $n=1.15$, $\qrms=18.2 \mu$K. In other words, although the total normalization has risen by a mere $3\%$, the slope estimate has risen by $12\%$ and the peak likelihood has been shifted to higher $n$ and lower $\qrms$. This indicates that the higher angular resolution data from the Tenerife experiment contains slightly more power on small scales. As explained in Section~\ref{5}, this is not unexpected, since the presence of a Doppler peak would cause a rise in the power spectrum at higher $l$.

The large angular scale CMB observations from Tenerife and COBE probe fluctuations that have yet to go non-linear and the shape of the power spectrum is thus insensitive to the exact abundance of the baryonic mass ($\Omega_b$) and to the value of $h=\ho/100\kmspmpc$. However, together the Tenerife and COBE observations provide a direct measure of the CMB power spectrum normalisation, against which one can compare intermediate scale observations and hence discriminate between competing cosmological models. Here we stress the importance in the agreement between the derived normalisation for the independent Tenerife and COBE experiments, which are subject to different systematic errors and different foreground contamination. The accuracy to which we know the normalisation of the power spectrum clearly becomes an issue when comparing with smaller scale observations to determine cosmological parameters.
This is a particular concern if, as has been suggested \cite{crittenden,steinhardt,abbott84}, a component of the large scale anisotropy may be due to tensor metric perturbations produced from a background of gravitational waves. These can arise naturally in inflationary scenarios and would contribute a component $\clt=<|a_{lm}^{T}|^2>$ to the observed CMB angular power spectrum. 
The ratio of the tensor modes \clt to the scalar modes \cls (primordial density fluctuations) is highly suppressed for fluctuations contained within the horizon volume at recombination and hence their contribution is only significant on scales \simgt $2\dg$. Consequently the existence of a tensor contribution has implications when comparing the large scale anisotropy level with that on smaller scales and with large scale structure observations in order to test cosmological models.  The anisotropy measurements from Tenerife and COBE fix the sum of
\clt and \cls, but the separation of the two terms requires a comparison with smaller scale observations under the  assumption that a given cosmological model is correct \cite{crittenden,steinhardt}. In the case of power law inflation a relation exists between the tensor to scalar ratio $C_2^{(T)}/C_2^{(S)}$ and the slope of the primordial power spectrum \cite{crittenden,steinhardt}: 
\be
C_2^{(T)}/C_2^{(S)} \approx 7(1-n)
\label{eq:tensors}
\ee
from which we see that the two contributions are comparable at $n=0.85$ with \clt decreasing relative to \cls for higher values of $n$. Thus the limit of $n \ge 1.0$ obtained from the analysis of the combined Tenerife and COBE data in Section~\ref{likejoint} implies that it is unlikely that the tensor component will be dominant in such models.
This is investigated in more detail by comparing with medium-scale anisotropy results.

The work required to place the existing medium scale anisotropy results  into a common statistical framework and then to compare them with the predictions of cosmological models, is carried out in Sections~\ref{comp1} and \ref{comp2}.

\subsection{Conclusions}

We have presented a summary analysis of the measurements taken at Dec=+40\degg~by the Tenerife experiments at 10, 15 and 33 GHz. After accounting for both local atmospheric and discrete radio source foregrounds, the 15 and 33 GHz data at high Galactic latitude are seen to contain statistically significant signals which have their origin in common hot and cold features. The cross-correlation function between the data at these two frequencies demonstrates that the amplitudes and shapes of the structures detected at 15 and 33 GHz are similar (paper II). This, combined with our measurements at the lower frequency of 10 GHz, implies that the CMB signal dominates over the Galactic contribution at 15 GHz, and that the maximum possible Galactic contribution at 33 GHz is smaller than 10 \% of the detected signal. Our best estimate of the cosmological signal is $\qrms=22^{+10}_{-6} \mu$K for an $n=1$ inflationary spectrum. This amplitude is reduced by $4\mu$K over that previously reported for the same data set and results from an improved separation of signal from atmospheric noise. Comparison of the Tenerife and COBE two year anisotropy detections by means of the likelihood function allows a detailed investigation of the allowed parameter space for a power law model of the fluctuation spectrum. The best fit values of $n$ and \qrms are $1.37$ and $16 \mu$K, and marginalising over \qrms we find both data sets consistent with $n$ in the range $1.0\le n \le 1.6$. These results support inflationary models, which predict $n \simeq 1$ and future improvements in the Tenerife data \cite{gutierrez97} and use of the 4 year COBE data will narrow the range of allowed $n$. Improvements of this kind are important to determine the power spectrum normalisation, since {\em only} on these large angular scales is it possible to place limits on the tensor to scalar ratio independent of the precise details of the cosmological model. In Sections~\ref{comp1},\ref{comp2} improved limits have been placed on $n$, by combining the large scale anisotropy measurements from Tenerife and COBE with observations of medium scale anisotropy measurements, 
but at the expense of assuming an underlying cosmological model (in our case CDM).

\section[Constraints-I]{Constraints on the Cosmological Parameters using current CMB observations-I}
\label{comp1}
 
Observations of the Cosmic Microwave Background (CMB) radiation provide information about epochs and physical scales that are inaccessible to conventional astronomy. In contrast to traditional methods of determining cosmological parameters, which rely on the combination of results from local observations \cite{os}, CMB observations provide direct measurements \cite{be87,white} over cosmological scales, thereby avoiding the systematic uncertainties and biases associated with conventional techniques. As mentioned in Section~2 the principal cosmological information is contained in the acoustic peaks \cite{hu,scott} in the power spectrum, which are generated during acoustic oscillations of the photon-baryon fluid at recombination \cite{cdm2}. 
The main acoustic peak, known as the `Doppler peak', is a strong prediction of contemporary cosmological models with adiabatic fluctuations and is expected to occur on an angular scale $\sim 1\dg$ (see Section~2); its observation is a major goal of observational cosmology. 
In the case that this peak was not observed, this would imply either that medium-scale primordial CMB fluctuations had been wiped out by reionization \cite{cdm2} or that there is a fundamental flaw in our theory. On the contrary, a conclusive observation of the first peak would provide strong support for current theoretical models and the determination of its angular position would constitute a direct probe of the large scale geometry of the Universe.  
The angular scale $l_p$ of the main peak reflects the size of the horizon at last scattering of the CMB photons and thus depends almost entirely \cite{hu,kamio} on the total density of the Universe according to $l_p \propto 1/\sqrt{\Omega}$.  
In conventional inflationary theory \cite{guth}, one expects the Universe to be flat with $\Omega=1.0$, which can be achieved if the total mass density is equivalent to the critical density or if there is a contribution from a cosmological constant $\Lambda$.  
The height of the peak provides additional cosmological information since it is directly proportional to the fractional mass in baryons $\Omega_b$ and also varies according to the expansion rate of the Universe as specified by the Hubble constant $\mbox{H}_0$; in general \cite{hu} for baryon fractions $\Omega_b \simlt 0.05$, increasing \ho reduces the peak height whilst the converse is true at higher baryon densities. Furthermore, by measuring the amplitude of the intermediate scale CMB fluctuations relative to those on large scales it is possible to place tight limits on the spectral slope $n$ of the initial primordial spectrum of fluctuations. The latter is predicted by inflationary theory to be approximately scale invariant, in which case $n \simeq 1.0$, although the presence of a background of primordial gravity waves \cite{crittenden,steinhardt} would generally lead to lower values of $n$. 
Thus, in summary, by comparing large and intermediate scale CMB observations and tracing out the Doppler peak, it is possible to directly estimate $\Omega$, $\Omega_b$ and \ho and to probe inflationary theory and the existence of primordial gravity waves. 
Recent improvements in the quality of CMB data, in particular on the angular scales probed by the CAT and Saskatoon experiments, now make this exercise of great interest.

Here we present the analysis and results published in Hancock \et 1997 \cite{me96}.
In Section~\ref{meth} we describe the method applied for the intercomparison of experiments and models. In Section~\ref{res} we proceed to the discussion of the results obtained.
For more information see also \cite{me96,moriondh,ppeuc,thesis}.

\subsection{Method}
\label{meth}
 
Clear detections of CMB anisotropy have been reported by a number of different groups, including the COBE satellite \cite{smoot,bennett}; ground-based switching experiments such as Tenerife \cite{me96,nature94}, Python \cite{python} South Pole \cite{spole} and Saskatoon \cite{sask}; balloon mounted instruments such as ARGO \cite{argo1}, MAX \cite{max}, and MSAM \cite{msam1,msam2} and more recently the ground-based interferometer CAT \cite{cat} (these were the most recent detections at the time of publication of paper Hancock \et 1997 (\cite{me97})).
Recently other new detections have been reported and are considered and included in the analysis carried out in Section~\ref{comp2}. 
Given the difficulties inherent in observing CMB anisotropy, it is possible that some of these results are contaminated by foreground effects and it is clear that determining the form of the CMB power spectrum in order to trace out the Doppler peak requires a careful, in-depth consideration of the CMB measurements from the different experiments within a common framework. 

\begin{figure}
\centerline{\psfig{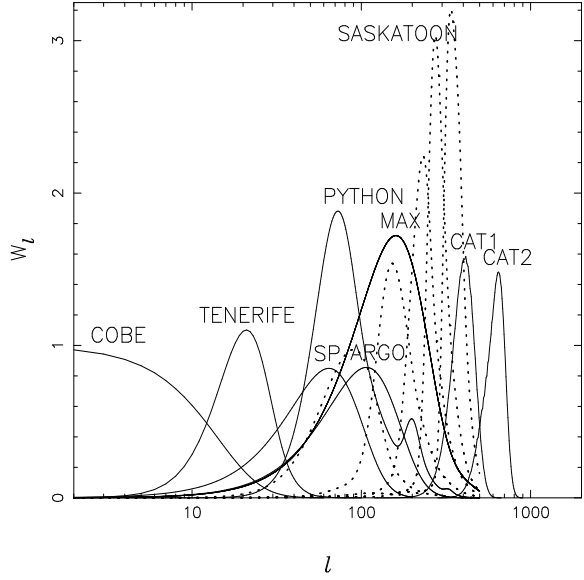}}
\caption{The window functions for the experiments listed in Table~\ref{tabdat} \label{fig:windows}}
\end{figure}

We consider all of the latest CMB measurements at the time of the analysis described in paper Hancock \et (\cite{me97}), including results from COBE, Tenerife, MAX, Saskatoon and CAT, with the exception of the MSAM results (see below) and the MAX detection in the Mu Pegasi region which is contaminated by dust emission \cite{fischer}. (In Section~\ref{comp2} this data set is updated by inclusion of some more recent detections and a comparison of the results is there given.)
The competing models for the origin and evolution of structure predict \cite{be87,hu}, the shape and amplitude of the CMB power spectrum and its Fourier equivalent, the autocorrelation function $C(\theta) =<\Delta T({\bf n_1})\Delta T({\bf n_2})>$ where ${\bf n_1} \cdot {\bf n_2} =\cos \theta$. 
According to Section~\ref{basic} expanding the intrinsic angular correlation function $C(\theta)$ in terms of spherical harmonics one obtains 
$C(\theta)= \sum_{l\ge2}^{\infty} (2l+1) C_l P_l(\cos\theta) /4 \pi$,
where low order multipoles $l$ correspond to large angular scales $\theta$ and large $l$-modes are equivalent to small angles on the sky.  The $C_l$'s are predicted by the cosmological theories and contain all of the relevant statistical information for models described by Gaussian random fields \cite{be87}. As mentioned in Section~\ref{basic} the different experiments sample different angular scales according to their {\em window functions} $W_l$ \cite{whitewindow1,whitewindow2}. The window function $W_l$ specifies the relative sensitivity of an experiment to a given $l$-mode, and the observed power in CMB fluctuations as seen through a window $W_l$ is given by 
\be 
C_{obs}(0)=\left (\frac{\Delta T_{obs}}{T} \right )^2 = \sum_{l \ge 2}^{ \infty} (2l+1) C_l W_l/4 \pi.
\label{eq:cobs}
\ee 
Given $W_l$, then for the $C_l$'s corresponding to the theoretical model under consideration it is possible to obtain the value of $\Delta T_{obs}$ one would expect to observe using the chosen experiment (see also Section~\ref{basic}).  This value can then be compared to the value actually observed to test the cosmological model.  Shown in Fig.~\ref{fig:windows} are the window functions for the various configurations of the experiments considered.
 
On the largest scales corresponding to small $l$, new COBE \cite{bennett} and Tenerife \cite{me96} results improve the power spectrum normalisation, whilst significant gains in knowledge at high $l$ are provided by new results from the Saskatoon and CAT experiments. The full data set spans a range of $2$ to $\sim 700$ in $l$, sufficient to test for the main Doppler peak out to $\Omega =0.1$. We take the reported CMB detections and convert them to a common framework of flat bandpower results \cite{bond1,bond2} as given in Table~\ref{tabdat}. This is carried out as follows.

\begin{figure}
\centerline{\psfig{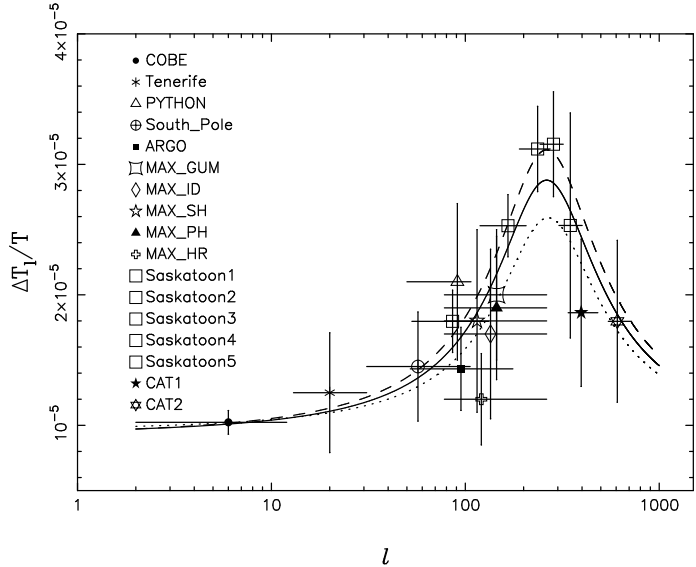}}
\caption{The data points from Table~\ref{tabdat} are shown compared to the best fit analytical CDM model. The dotted and dashed lines show the best fit models which are obtained when the Saskatoon calibration is adjusted by $\pm 14\%$. The data points from the MAX experiment are shown offset in $l$ for clarity \label{fig:fig2a}}
\end{figure}

\begin{table}
\begin{center}
\begin{minipage}{4.5in}
\begin{tabular}{|l|c|c|c|c|c|c|} \hline
Experiment & $\Delta T_l$ ($\mu$K) & $\sigma$ ($\mu$K) & $l_e$ & $l_l$  &$l_u$
& Reference \\ 
COBE       &     27.9 &     2.5 &   6 &   2 &   12 & \cite{bennett}\\
Tenerife   &     34.1 &     12.5 &   20 &   13 &   31 & \cite{me96}\\
PYTHON     &     57.2 &     16.4 &   91 &   50 &   107 & \cite{python}\\
South Pole &     39.5 &     11.4 &   57 &   31 &   106 & \cite{spole}\\
ARGO       &     39.1 &     8.7 &   95 &   52 &   176 & \cite{argo1}\\
MAX GUM    &     54.5 &     13.6 &   145 &   78 &   263 & \cite{max}\\
MAX ID     &     46.3 &     17.7 &   145 &   78 &   263 & ``\\
MAX SH     &     49.1 &     19.1 &   145 &   78 &   263 & ``\\
MAX PH     &     51.8 &     15.0  &   145 &   78 &   263 & ``\\
MAX HR     &     32.7 &     9.5 &   145 &   78 &   263 & ``\\
Saskatoon1 &     49.0 &     6.5 &   86 &   53 &   132  & \cite{sask}\\
Saskatoon2 &     69.0 &     6.5 &   166 &   119 &   206 & ``\\
Saskatoon3 &     85.0 &     8.9 &   236 &   190 &   274 & ``\\
Saskatoon4 &     86.0 &     11.0 &   285 &   243 &   320 & ``\\
Saskatoon5 &     69.0 &     23.5 &   348 &   304 &   401 & ``\\
CAT1       &     50.8 &     15.4 &   396 &   339 &   483 & \cite{cat}\\
CAT2       &     49.0 &     16.9 &   608 &   546 &   722 & ``\\
\hline
\end{tabular}
\caption{Details of data results used} 
\label{tabdat} 
\end{minipage}
\end{center}
\end{table}
The CMB anisotropy measurements are converted to bandpower estimates $\Delta T_l \pm \sigma$ assuming in each case a flat spectrum of $C_l$ centred on the effective multipole $l_e$ (see below) of the window function. $l_l$ and $l_u$ represent the lower and upper points at which the window of each configuration reaches half of its peak value.  In order to use the observed anisotropy levels to place constraints on the CMB power spectrum one must in general know the form of the $C_l$ under test.  However, in most cases the form of $C_l$ can be represented by a flat spectrum $C_l \propto C_2/(l(l+1))$ over the width of a given experimental window, so that the bandpower is $\Delta T_l/T=\sqrt{C_{obs}(0)/I(W_l)}$, where we define $I(W_l)$ according to Bond \cite{bond1,bond2} as $I(W_l)=\sum_{l=2}^{\infty} (l+0.5)W_l/(l(l+1)$. This bandpower estimate is centred on the effective multipole $l_e=I(lW_l)/I(W_l)$.  In many instances experimenters now report results directly for a flat spectrum and when this is not so we have converted the quoted power in fluctuations into the equivalent flat band estimate.  Each group has obtained limits on the intrinsic anisotropy level using a likelihood analysis 
(see \eg Hancock \ea\ \cite{nature94}), which incorporates uncertainties due to random errors, sampling variance \cite{samvariance} and cosmic variance \cite{cosvariance2,cosvariance1} (see Section~\ref{basic} and \cite{thesis}).  The errors in $\Delta T_l$ quoted in column 3 of Table~\ref{tabdat} are at 68 \% confidence and have been obtained by averaging the difference in the reported 68\% upper and lower limits and the best fit $\Delta T_l$. Since the form of the likelihood function is in general only an approximation to a Gaussian distribution this averaging introduces a small bias into the results (see Section~\ref{comp2}, \cite{gr1}).  Results from the MSAM experiment are not included here, because they do not provide an independent measure of the power spectrum since their angular sensitivity and sky coverage are already incorporated within the Saskatoon measurements. (If the correlation is small, as recent information seems to indicate, then this experiment may be included in the analysis; this is done in Section~\ref{comp2}.) 
Netterfield \ea\ \cite{sask} report good agreement between the MSAM double difference results and Saskatoon measurements, although the discrepancy with the MSAM single difference data is yet to be resolved.
The data points from Table~\ref{tabdat} are plotted in Figure~\ref{fig:fig2a},
in which the horizontal bars represent the range of $l$ contributing to each data point. There is a noticeable rise in the observed power spectrum at $l\simeq 200$, followed by a fall at higher $l$, tracing out a clearly defined peak in the spectrum.  In the past several groups \cite{scott,kami,ratra,Ganga96} have attempted to determine the presence of a Doppler peak, but only now are the data sufficient to make a first detection and to put constraints on the closure parameter $\Omega$.  As a first step, we adopt a simple three parameter model of the power spectrum, which we find adequately accounts for the properties of the principal Doppler peak for both standard Cold Dark Matter (CDM)  models \cite{cdm1,cdm2} and open Universe ($\Omega<1$) models \cite{kami}. The functional form chosen is a modified version of that used in Scott, Silk \& White 
\cite{scott} --- we choose the following: 
\be
l(l+1)C_l=6C_2\left(1+\frac{A_{peak}}{1+y(l)^2}\right ) {\huge /}
\left(1+\frac{A_{peak}}{1+y(2)^2}\right )
\label{eq:peak}
\ee 
where $y(l)=(\log_{10}l-log_{10}(220/\sqrt{\Omega}))/0.266$. In this representation $C_2$ specifies the power spectrum normalisation, whilst the first Doppler peak has height $A_{peak}$ above $C_2$, width $\log_{10}l=0.266$ and for $\Omega=1.0$ is centred at $l\simeq 220$.  By appropriately specifying the parameters $C_2$, $A_{peak}$ and $\Omega$ it is possible to reproduce to a good approximation the $C_l$ spectra corresponding to standard models of structure formation with different values of $\Omega$, $\Omega_b$ and $\mbox{H}_0$. Such a form will not reproduce the structure of the {\em secondary\/} Doppler peaks, but we have checked the model against the overall form of the $\Omega =1$ models of Efstathiou and the open models reported in Kamionkowsky \ea\ 
\cite{kami} and find that this form adequately reflects the properties of the main peak. This satisfies our present considerations since the current CMB data are not yet up to the task of discriminating the secondary peaks.  Varying the three model parameters in equation~(\ref{eq:peak}) we form $C_l$ spectra corresponding to a range of cosmological models, which are then used in equation~(\ref{eq:cobs}) to obtain a simulated observation for the $i$th experiment, before converting to the bandpower equivalent result $\Delta T_l[C_2,A_{peak},\Omega](i)$.  The chi-squared for this set of parameters is given by
\begin{displaymath}
\chi^2(C_2,A_{peak},\Omega)=\sum_{i=1}^{nd}
\frac{(\Delta T_l^{obs}(i) -\Delta T_l[C_2,A_{peak},
\Omega](i))^2}{\sigma_i^2}, 
\label{eq:chi}
\end{displaymath}
for the $nd$ data points in Table~\ref{tabdat} and the relative likelihood function is formed according to $L(C_2,A_{peak},\Omega) \propto \exp( - \chi^2(C_2,A_{peak},\Omega)/2)$.  We vary the power spectrum normalisation $C_2$ within the 95 \% limits for the COBE 4-year data \cite{bennett} and consider $A_{peak}$ in the range 0 to 30 and values of the density parameter up to $\Omega=5$.  The data included in the fit are those from Table~\ref{tabdat}, which with the exception of Saskatoon include uncertainties in the overall calibration.
There is a $\pm 14$\% calibration error in the Saskatoon data, but since the Saskatoon points are not independent this will apply equally to all five points \cite{sask}.  The likelihood function is evaluated for three cases: (i) that the calibration is correct, (ii) the calibration is the lowest allowed value and (iii) the calibration is the maximum allowed value.  In each case the likelihood function is marginalised over $C_2$ before calculating limits on the remaining two parameters according to Bayesian integration with a uniform prior.

\subsection{Results and Discussion}
\label{res}

\begin{figure}
\hspace*{-0.5in}
\vbox{%
\hspace*{0.5in}
\psfig{figure=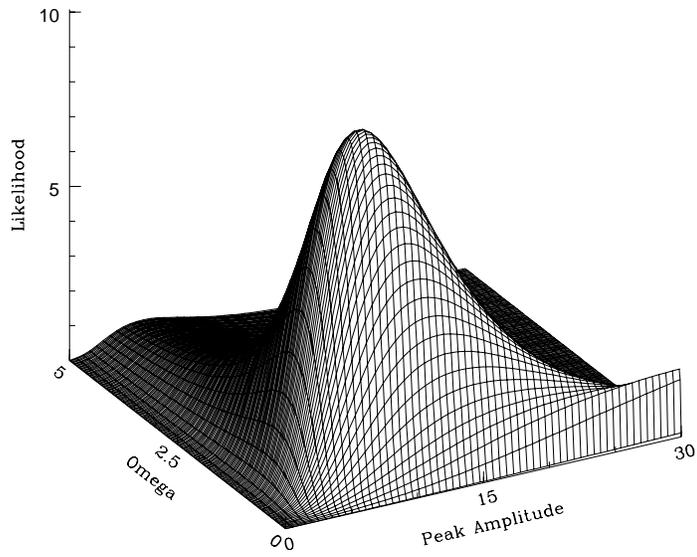,height=4.2in,width=4.2in,angle=0}
\hspace*{1.3in}
\psfig{figure=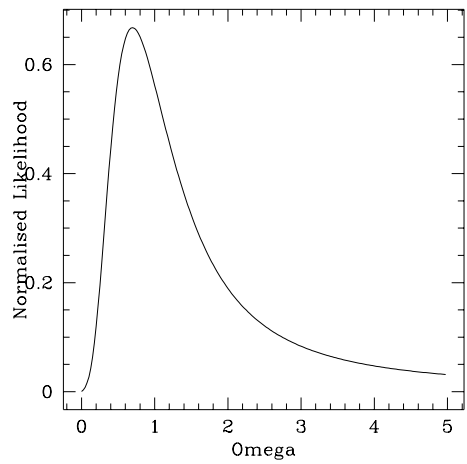,height=2.5in,width=2.5in,angle=0}}
\caption{Top Figure: The likelihood surface for $\Omega$ and $A_{peak}$. (The nominal Saskatoon calibration is assumed.
Bottom Figure: The 1-D conditional likelihood curve for $\Omega$.}
\label{fig:3d}
\end{figure}

In Fig.~\ref{fig:3d} the likelihood function obtained from fitting the model $C_l$ spectra to the data of Table~\ref{tabdat} is shown plotted as a function of the amplitude and position ($\Omega$) of the Doppler peak. The highly peaked nature of the likelihood function in Fig.~\ref{fig:3d} is good evidence for the presence of a Doppler peak localised in both position ($\Omega$) and amplitude. In Fig.~\ref{fig:3d}
we also show the 1-D conditional likelihood curve for $\Omega$, obtained by cutting through the surface shown in the left hand side of Fig.~\ref{fig:3d} at the best-fit value of $A_{peak}$. The best fit value of $\Omega$ is 0.7 with an allowed 68\% range of $0.30 \le \Omega \le 1.73$.
In Figure~\ref{fig:fig2a} the best fit model, represented by the solid line, is shown compared to the data points, assuming no error in the calibration of the Saskatoon observations. The chi-squared per degree of freedom for this model is $0.9$, implying a good fit to the data.  The peak lies at $l=263_{-94}^{+139}$ corresponding to a density parameter $\Omega =0.70_{-0.4}^{+1.0}$; the height of the peak is $A_{peak}=9.0_{-2.5}^{+4.5}$.  The dashed and dotted lines show the best fit models ($\Omega =0.70_{-0.37}^{+0.92}$, $A_{peak}=11.0_{-4.0}^{+5.0}$ and $\Omega =0.68_{-0.4}^{+1.2}$, $A_{peak}=6.5_{-2.0}^{+3.5}$ respectively) assuming that the Saskatoon observations lie at the upper and lower end of the permitted range in calibration error.  
These likelihood results using the analytic form for the $C_l$ and the results from a chi-squared goodness of fit analysis using exact models (see below) imply that independent of calibration uncertainties in the data, current CMB data are inconsistent with cosmological models with $\Omega \simlt 0.3$.

The analytic approximation to the true $C_l$ such as we use here, is a useful general tool, but as a detailed check we have also applied the chi-squared goodness of fit test to actual COBE normalised $C_l$ models.
A detailed comparison of the CMB data with these models is considered in Section~\ref{comp2}. For example, considering a range of CDM models with varying $\Omega$, in order to find the lowest $\Omega$ compatible with the observations, we have considered exact models with $\ho=50 \kmspmpc$ and $\Omega_b=0.03$ for $\Omega =0.1 - 0.5$ \cite{kami}.  We find that $\Omega=0.5$ is allowed, $\Omega=0.3$ and below are completely ruled out (95\% confidence) and $\Omega=0.4$ is excluded unless all the Saskatoon points have the minimum allowed calibration. This result is confirmed in Section~\ref{comp2} for a more complet set of open models.

\subsection{Conclusions}

Our current results provide good evidence for the Doppler peak, verifying a crucial prediction of cosmological models and providing an interesting new measurement of fundamental cosmological parameters.
This first estimate of the angular position of the Doppler peak is used to place a new direct limit on the curvature of the Universe, corresponding to a density of $\Omega=0.7^{+1.0}_{-0.4}$, consistent with a
flat Universe.  Very low density `open' Universe models are inconsistent with this limit unless there is a significant contribution from a cosmological constant.
In section~\ref{comp2} a detailed comparison of the CMB data is made with the theoretical power spectra predicted by a range of flat, tilted, open models and models with non-zero cosmological constant.


\section[Constraints-II]{Constraints on the Cosmological Parameters using current CMB observations-II}
\label{comp2}

In this Section we give a detailed comparison of the CMB data with the theoretical power spectra predicted by a range of flat, tilted, open models and models with non-zero cosmological constant. 
In this analysis we only consider scalar fluctuations and we do not include reionization. This is an extension of the work carried out in Section~\ref{comp1} in which we used an analytic approximation of current CMB models. 
Here we consider some of the exact fourth-year COBE normalized angular power spectrum with fourth-year COBE normalization. The precise form of the Doppler peaks depend on the nature of the dark matter, and the values of $\Omega$, \ob, and \ho. Thus in order to use the medium-scale anisotropy results e.g. to provide additional leverage on the spectral index determination, it is necessary to adopt a given cosmological model. 
Each Section gives the results for each set of models of structure formation used in this intercomparison of data and models.
For more information see also \cite{me96,moriondh,moriondr,ppeuc,thesis}.
Other related analysis usig a $\chi^{2}$ technique have been developped by, among others, \cite{Ganga96}, \cite{line}, etc.

Ganga et al. (1997a, hereafter GRGS)) developed a technique to account for uncertainties, such as those in the beamwidth and the calibration, in likelihood analyses of CMB anisotropy data. 
This technique has been used in conjunction with theoretically-predicted CMB 
anisotropy spectra in analyses of the Gundersen et al. (1995) UCSB South Pole 1994 data (SP94), the Church et al. (1997) SuZIE data, the MAX 4+5 data (Tanaka et al. 1996; Lim et al. 1996), the Tucker et al. (1993) White Dish data, and the de Bernardis et al. (1994) ARGO data (GRGS; Ganga et al. 1997b, 1998; Ratra et al. 1998, 1999a, hereafter R99a). A combined analysis of all these data sets is presented in Ratra et al. (1999b, hereafter R99b) in which a joint likelihood analysis of these data sets is carried out.
Bond \& Jaffe (1997) have also analysed the SP94 data and the Saskatoon (Netterfield \et 1997) data.  
A similar analysis of CMB anisotropy data from the Python I, II, and III observations performed at the South Pole (Dragovan et al. 1994, hereafter D94; Ruhl et al. 1995b, hereafter R95; Platt et al. 1997) 
is described in \cite{me99}.
The last Section gives a brief account of these other results in conjunction with the ones obtained here. 

\subsection{Data}

\begin{figure}
\begin{center}
\psfig{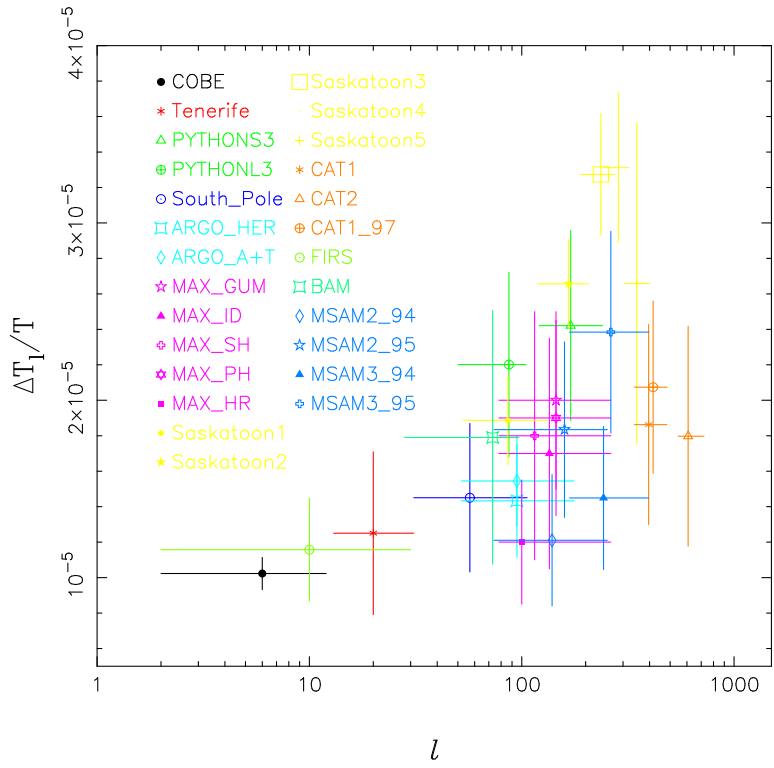}
\caption{The updated data set (c). \label{fig:datup2}}
\end{center}
\end{figure}

We apply a fit procedure to three sets of CMB data first: the CMB data given in Table~\ref{tabdat} of Section~\ref{comp1}, where the COBE data consists in one multipole for COBE (namely $C_{2}$, obtained from the value of the \qrms=18 \uk\ for the 4-year COBE data \cite{bennett}) which we refer to as case (a); 
the same as the data of Table~\ref{tabdat} apart from COBE where we consider 4 of the 8 multipole bands of the angular power spectrum extracted from the 4-year COBE data \cite{tegmark}, which we refer to as case (b), and the data consisting basically in the extension of data set (a) to include (1) the CAT1-97 point at $\ell = 422$ \cite{cat2}; (2) the inclusion of new points
from experiments Python (Python III, Platt \etal\ 1997), 
MSAM (the 2nd and 3rd flights, Cheng \etal\ 1996, 1997, Ratra
\etal\ 1997), ARGO (Aries+Taurus region, Masi \etal\ 1996),
FIRS \cite{firs} and BAM \cite{bam}, and with the latest calibration
correction to the Saskatoon data (i.e. increased by 5\%, Leitch, 
private communication); which we refer to as case (c) and is displayed in Table~\ref{datup2} and shown in Fig~\ref{fig:datup2}.
In the case of data set (b), in order to avoid eventual correlations between these data points, we have applied the fitting analysis to the 4 even or 4 odd multipole bands in conjuntion with the remaining data points. The results are similar whether we use the even or the odd multipoles. The use of 4 instead of 1 multipole is particularly useful in discriminating the value of the spectral index $n$.
\begin{table}
\begin{center}
\caption{Details of data results (c) used}
\begin{tabular}{|l|c|c|c|c|c|c|} \hline
Experiment & $\Delta T_l$ ($\mu$K) & $\sigma$ ($\mu$K) & $l_e$ & $l_l$  &$l_u$ & Ref
\\ \hline
COBE         & 27.9   & 2.5    &6   &2   &12  & \cite{bennett} \\
Tenerife     & 34.1   & 12.5   &20  &13  &31  & \cite{me96} \\
PYTHONS3     & 66.0   & 14.6   &170 &121 &240 & \cite{Platt} \\ 
PYTHONL3     & 60.0   & 14.2   &87  &50  &105 & `` \\
South Pole   & 39.5   & 11.4   &57  &31  &106 & \cite{spole} \\
ARGO HER     & 39.1   & 8.7    &95  &52  &176 & \cite{argo1} \\
ARGO A+T     & 42.1   & 6.9    &95  &52  &176 & \cite{argo}\\ 
MAX GUM      & 54.5   & 13.6   &145 &78  &263 & \cite{max} \\
MAX ID       & 46.3   & 17.7   &145 &78  &263 & `` \\
MAX SH       & 49.1   & 19.1   &145 &78  &263 & `` \\ 
MAX PH       & 51.8   & 15.0   &145 &78  &263 & `` \\
MAX HR       & 32.7   & 9.5    &145 &78  &263 & `` \\ 
Saskatoon1   & 51.4   & 6.8    &86  &53  &132 & \cite{sask} \\
Saskatoon2   & 72.4   & 6.8    &166 &119 &206 & `` \\
Saskatoon3   & 89.2   & 9.4    &236 &190 &274 & `` \\
Saskatoon4   & 90.3   & 11.5   &285 &243 &320 & `` \\
Saskatoon5   & 72.5   & 24.7   &348 &304 &401 & `` \\
CAT1         & 50.8   & 15.4   &396 &339 &483 & \cite{cat} \\
CAT2         & 49.0   & 16.9   &608 &546 &722 & ``\\
CAT1 97      & 56.5   & 13.2   &396 &339 &483 & \cite{cat2} \\
FIRS         & 31.6   & 7.9    &10  &2   &30  & \cite{firs} \\
BAM          & 48.8   & 19.5   &73  &28  &97  & \cite{bam} \\
MSAM2 94     & 33.0   &10.1    &159 &74  &253 & \cite{msam2,msam95,ratra} \\
MSAM2 95     & 50.0   & 13.5   &159 &74  &253 & `` \\
MSAM3 94     & 39.5   & 11.0   &263 &168 &397 & `` \\
MSAM3 95     & 65.0   & 15.5   &263 &168 &397 & `` \\
\hline
\end{tabular}
\label{datup2}
\end{center}
\end{table}
\newline
\indent
As already mentioned in Section~\ref{comp1} the different experiments sample different angular scales according to their window functions, $W_{l}$, \cite{whitewindow1,whitewindow2}. 
The 4 window functions corresponding to the COBE experiment were obtained considering Gaussian functions centered in $\langle l \rangle$ with dispersion $(\Delta l)^{2}$, where the parameters are given in Tegmark 1996 \cite{tegmark}.
\newline
\indent
The conversion of data to a common framework in terms of flat bandpower estimates \cite{bond1,bond2} follows the same procedure as in Section~\ref{comp1}. In Table~\ref{datup2} we display these bandpower estimates $\Delta T_{l} \pm \sigma$ where as before $l_{l}$ and $l_{u}$ represent the lower and upper points at which the window function for each configuration reaches half of its peak value. 
As in Section~\ref{comp1}, the errors in $\Delta T_{l}$ quoted in column 3 are 68\% confidence limits and have been obtained by averaging the difference in the reported 68\% upper and lower limits and the best fit $\Delta T_{l}$.


\subsection{Method}

We pursued the same method as described in Section~\ref{comp1}. 
The process by which the error bars quoted in Table~\ref{datup2} are computed introduces a small bias into the results since the likelihood function is in general only an approximation to a Gaussian distribution. 
In order to assess the significance of this we applied the fitting analysis to an alternative set of data which consisted of the mean value of the data and corresponding error bars. We find no significant difference in the results obtained. 
We consider that a given model offers an acceptable chi-squared fit when $P(\chi^{2}) \geq 0.05$. We increase the resolution of the parameters grid via interpolation procedures and compute the $\chi^{2}$ and corresponding $P(\chi^{2})$ for each model. 
Including the 95$\%$ COBE normalization uncertainty and increasing the resolution of the parameter grid using interpolation procedures, we computed the best-fit model, its chi-squared value, the corresponding probability and the likelihood function. 
For each parameter both the conditional and marginal distributions were obtained as corresponding confidence intervals.


\subsection{Scale invariant flat CDM models}

\begin{figure}
\hspace*{1in}
\psfig{figure=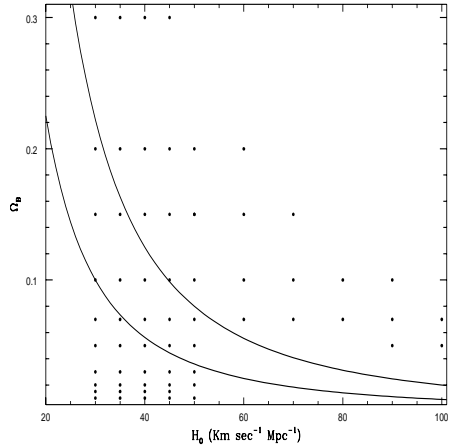,height=2.5in,width=2.5in,angle=-90}
\caption{The COBE normalised CDM models with acceptable chi-squared fits to the CMB data (for case (i) calibration) are plotted as dots in the $\Omega_{b}-H_{0}$ space. Overlayed are the constraints on $\Omega_{b}h^{2}$ imposed by nucleosynthesis. \label{fig:omh2}}
\end{figure}

This work was initially applied to scale invariant ($n=1$), flat ($\Omega=1$) CDM models with $\Omega_{b}$=0.01-0.3 and $H_{0}$=30-100 km s$^{-1}$ Mpc$^{-1}$ (kindly provided by G. Efstathiou). We considered these models normalised to the 4-year COBE value of \qrms=18 \uk\ and applied the analysis to the data tabulated in Table~\ref{tabdat} of Section~\ref{comp1} (case (a)).
In Fig.~\ref{fig:omh2} we have considered these COBE normalised CDM models. A dot is placed in the appropriate place in the parameter space if the exact power spectrum corresponding to these parameters gives a fit to the data in Table~\ref{tabdat} of Section~\ref{comp1} with an acceptable $\chi^{2}$ value ($P(\chi^2) \ge 0.05$). A blank is left at that position if not.           
Overlying these power spectrum constraints is the limit $0.009 \leq \Omega_{b} h^{2} \leq 0.02 $ provided by nucleosynthesis of the light elements \cite{copi}.
As shown, the models offering an acceptable chi-squared fit ($P(\chi^2) \ge 0.05$) to the CMB power spectrum whilst simultaneously satisfying nucleosynthesis constraints \cite{copi}, are those with $0.05\le \Omega_b \le 0.2$, $30\kmspmpc \le \ho \le 50 \kmspmpc$. Considering the highest Saskatoon data calibration the constraints become $0.1 \leq \Omega_{b} \leq 0.2$, $30\kmspmpc \le \ho \le 35 \kmspmpc$. 
Allowing for the lowest Saskatoon data calibration relaxes the constraints up to $\ho=70\kmspmpc$ and $0.02 \leq \Omega_{b} \leq 0.2$
(see Hancock \ea \cite{me96}). 
In general, recent optical and Sunyaev-Zel'dovich observations of the Hubble constant \cite{ho3,ho1,ho2,lasenbyandhancock} imply \ho in the range $50-80 \kmspmpc$. Since the paper Hancock \ea \cite{me96} was first submitted, an alternative comparison of CMB data with models \cite{line} has appeared, which supports our conclusions that low values of \ho are favoured by the current CMB data.
%
\begin{table}
\begin{center}
\begin{tabular}{lcccc} \hline
                        &$H_{0}$    & $\Omega_{b}$    & $Q_{rms-ps}$         \\
                        &           &                 &                      \\
nBBN                    &           &                 &                      \\
Best Model                      & 30        & 0.02              &  17.64            \\
Conditional(1$\sigma$)          & 30-50     & 0.02-0.15            & 16.02-19.44              \\
Marginal(1$\sigma$)             & 30-60     &                      &                      \\
                                &           &                      &                       \\ 
                        &           &                 &                       \\
BBN                     &           &                 &                       \\
Best Model                      & 30        & 0.1                 &  16.38            \\
Conditional(1$\sigma$)          & 30-40     & 0.1-0.19            & 14.94-17.82              \\
Marginal(1$\sigma$)             & 30-50     & 0.04-0.15           & 16.20-19.98                    \\
\hline
\end{tabular}
\caption{Results of the fitting analysis considering: one multipole for 
COBE experiment (a), that Saskatoon calibration is correct (i) and 
with (BBN), without (nBBN), BBN constraints for scale invariant flat CDM models.}
\label{tabdois}
\end{center}
\end{table}
%
Alternatively we generated the angular power spectrum, \cls, for these models using a CMB code kindly provided by U. Seljak and M. Zaldarriaga. The grid considered has Hubble constant, $\ho$, values in the range $\ho=30-80 \kmspmpc $in steps of 5 $\kmspmpc$, baryon density, $\ob$, values in the range $\ob=0.02-0.22$ in steps of 0.01 and $\omeg=1$. 
We applied the same analysis to these models considering the data set (a), 
assuming that the Saskatoon calibration is correct (i). 
The results
are displayed in Table~\ref{tabdois}. 
In order to set constraints on the parameters of these set of models we consider the 68\% c.i. for the marginalized distribution of each parameter constrained to the BBN limits, these encompass $30 \kmspmpc \leq \ho \leq 50 \kmspmpc$, $0.04 \leq \ob \leq 0.15$ and 16.2 \uk $\leq$ \qrms $\leq$ 19.98 \uk.
It should be stressed that these results were obtained considering only the case (i) for the Saskatoon calibration. This analysis confirms the previous results obtained using a cruder method.


\subsection{Tilted flat CDM models}

\begin{figure}
\begin{center}
\hbox{%
\psfig{file=tilt-50-07.1.alt.ps,width=3in,angle=270}
\psfig{file=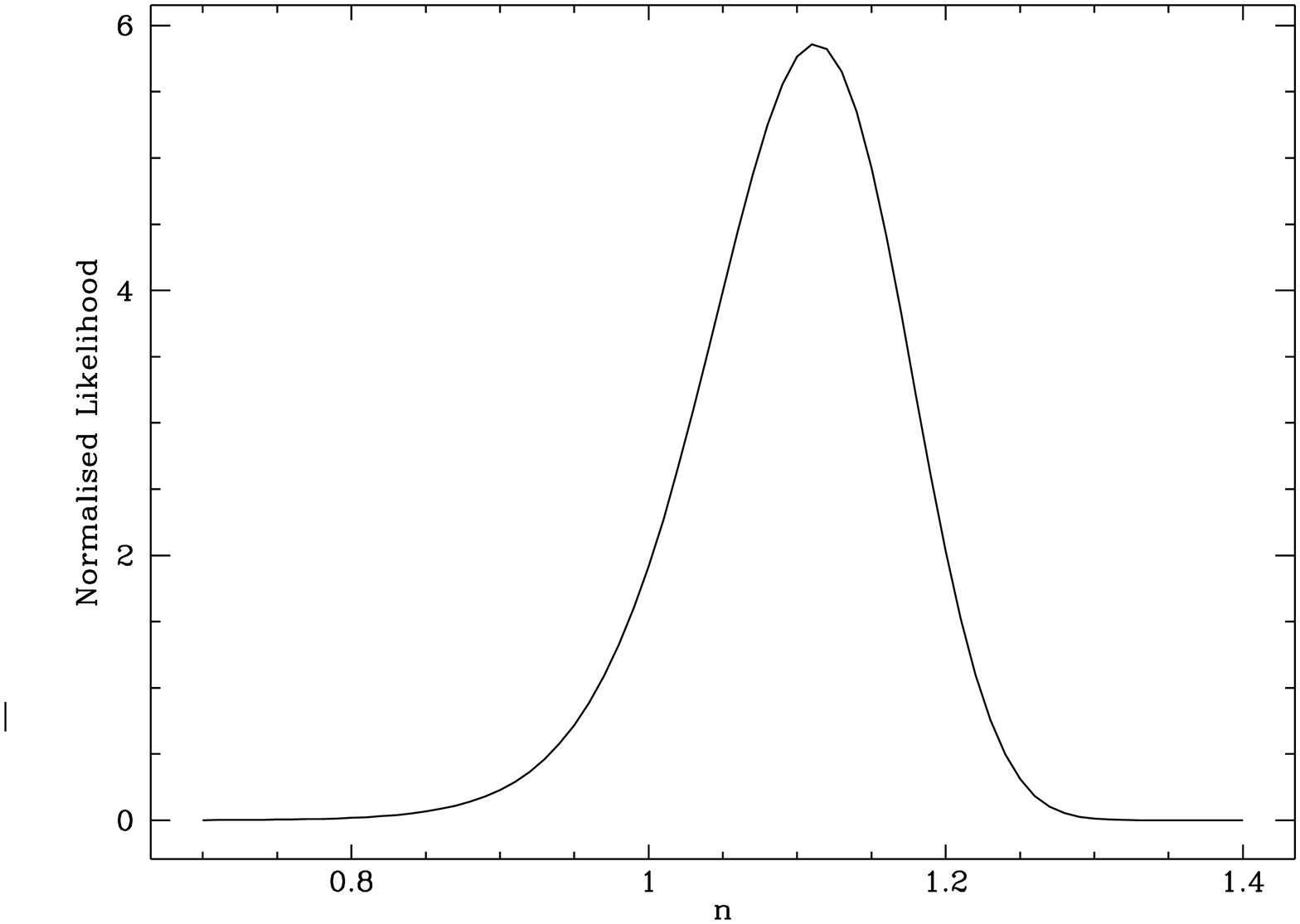,width=1.5in,height=2in,angle=270}}
\caption{ Tilted model with $h_{0}=0.5$, $\Omega_{b}=0.07$, $n=1.1$ and its 68 \%confidence interval (Left hand side Figure), likelihood function for the parameter $n$ (Right hand side Figure). \label {fig:tilt1}} 
\end{center}
\end{figure}

Initially we considered some of the best candidates for flat CDM models: 
(1) $h_{0}=0.3$, $\Omega_{b}=0.2$, 
(2) $h_{0}=0.45$, $\Omega_{b}=0.1$ and 
(3) $h_{0}=0.5$, $\Omega_{b}=0.07$ and allowed for the tilting of 
the power spectrum to vary within $n=0.7-1.4$. 
The fitting for the spectral index $n$ of the primordial fluctuations 
was done considering the two sets of data (a) and (b) mentioned above while at the same time allowing for the 2$\sigma$ COBE normalization uncertainty. 
The results are displayed in Table~\ref{3models}.
The third and fourth column display the uncertainties obtained considering the three calibration cases and taking extreme limits.
Comparing e.g. first and second column we conclude that in general the effect of using 4 multipoles for COBE data produces an increase of $n$.
\begin{table}
\begin{center}
\begin{tabular}{lcccc} \hline
\em{model}&             \multicolumn{4}{c}{$n$ (\em{$68\%$ \rm{conf. interval}})} \\ \hline
\vspace{5pt}
            &   (a),(i)&        (b),(i)&        (a),(i,ii,iii)&         (b),(i,ii,iii)                   \\     
$h_{0}=0.3$ &           $0.94 \pm 0.06$&        $0.99 \pm 0.07$&         $0.94 \pm 0.1$&        $0.99 ^{+0.11}_{-0.13}$ \\
\vspace{5pt}    
$\Omega_{b}=0.2$ &&&&   \\
$h_{0}=0.45$ &          $1.03 \pm 0.07$&        $1.08 \pm 0.07$&        $1.03 \pm 0.1$& $1.08 \pm 0.1$  \\
\vspace{5pt}
$\Omega_{b}=0.1$ &&&&   \\
$h_{0}=0.5$ &           $1.08 \pm 0.07$&        $1.11 \pm 0.07$&        $1.08^{+0.09}_{-0.12}$& $1.11^{+0.09}_{-0.11}$ \\
$\Omega_{b}=0.07$ &&&&  \\ \hline
\end{tabular}
\caption{Results of the fitting analysis for tilted flat models for: (a) data with 1 COBE multipole, (b) data with 4 COBE multipoles; (i) Saskatoon calibration is correct and (ii) lowest (iii) higher Saskatoon calibration.}
\label{3models}
\end{center}
\end{table}
 In particular fixing $H_{0}=50 \kmspmpc$ we find $n=1.1 \pm 0.1$ (68 \% confidence interval) allowing for the three calibration cases and taking extreme limits.
This tight limit rules out a significant contribution from a gravity wave background for these models, in the case of power law inflation \cite{liddle}, but is consistent with the prediction of $n \simeq 1.0$ for scalar fluctuations generated by inflation. 
In  Fig.~\ref{fig:tilt1} on the left hand side plot we present the model (3) and its 68 \% confidence interval assuming the case (i) for the calibration. 
On the right hand side plot is displayed the likelihood function for the parameter $n$ for model (3) which peaks at $n=1.1$ showing a significant estimate of the spectral index parameter based on the actual CMB data.
For details about the normalization of these models see e.g \cite{thesis}.

We then considered the 4-year COBE normalized tilted flat CDM models. We generated the angular power spectrum, \cls, using a CMB code provided by U. Seljak and M. Zaldarriaga. 
The grid considered  has \ho\ values in the range $\ho=30-80$ $\kmspmpc$ in steps of 5 $\kmspmpc$, \ob\  values in the range $\ob=0.02-0.22$ in steps of 0.01 and tilt in the range $n=0.6-1.4$ in steps of 0.1 for $\Omega=1$.
The results for case (a) are displayed in Table~\ref{table4} with and without superimposed BBN constraints as well as for the updated data set (c).
%
%
\begin{table}
\begin{center}
\begin{tabular}{lcccc} \hline

                        &$H_{0}$    & $\Omega_{b}$   & $n$   & $Q_{rms-ps}$         \\
(a),nBBN                    &           &                &       &                     \\
Best Model               & 30        & 0.22           & 0.92      & 17.95               \\
Conditional(1$\sigma$)   & 30-55     & 0.10-0.22      & 0.81-1.0  & 16.18-20.12              \\
Marginal(1$\sigma$)      & 30-70     &                & 0.78-1.12 &                \\
                        &           &                &       &                 \\       
(a),BBN                     &           &                &       &                  \\
Best Model               & 30        & 0.22           & 0.92      & 17.95               \\
Conditional(1$\sigma$)   & 30-30     & 0.13-0.22      & 0.81-1.0  & 16.18-20.12              \\
Marginal(1$\sigma$)      & 30-55     &                & 0.85-1.18 &                \\
                        &           &                &             &                 \\
(c),BBN                 &           &                &             &                 \\
Best Model               & 30        & 0.22           & 0.88        & 18.87          \\
Conditional(1$\sigma$)   & 30-30     & 0.13-0.22      & 0.78-0.98   & 16.63-21.10   \\
Marginal(1$\sigma$)      & 30-55     &                & 0.80-1.15   &                 \\ 
(2$\sigma$)              & 30-80     &                & 0.65-1.3    &                 \\ 
\hline
\end{tabular}
\caption{Results of the fitting analysis considering: one multipole for 
COBE experiment (a) and the updated data set (c), that Saskatoon calibration is correct (i) and 
with (BBN), without (nBBN), BBN constraints for tilted flat CDM models.}
\label{table4}
\end{center}
\end{table}

Considering the BBN constraints we obtain for the marginalized distribution of the tilt a 68\% c.i. of $0.85 \leq n \leq 1.18$ while not including this constraint implies an $n$ in the range 0.78-1.12 (68 \%). Adding new data points to data set (a) to get data set (c) affects the results by decreasing the value of the tilt $n$ from 0.92 to 0.88. The marginal distribution suffers a slight variation with 
an interval 0.85-1.18 for case (a) and 0.80-1.15 for case (c).
Both results are consistent with the prediction of $n \simeq 1.0$ for scalar fluctuations generated by inflation. 

Fixing $\ho=50 \kmspmpc$ the results obtained are tabulated in Table~\ref{table5}.
%
\begin{table}
\begin{center}
\begin{tabular}{lccc} \hline
                            & $\Omega_{b}$   & $n$   & $Q_{rms-ps}$         \\
                            &                &       &                      \\
(a),nBBN                            &                &       &                      \\
Best Model                   & 0.22           & 0.91  & 17.68             \\
Conditional(1$\sigma$)       & 0.15-0.22      & 0.81-1.0 & 15.89-19.66              \\
Marginal(1$\sigma$)               &                & 0.84-1.16   &                \\
                             &                &             &                \\ 
(a),BBN                      &                &                 &                    \\
Best Model                   & 0.08           & 1.07            & 16.79                 \\
Conditional(1$\sigma$)       & 0.05-0.08      & 0.98-1.18       & 15.14-18.82              \\
Marginal(1$\sigma$)          &                & 0.93-1.21       &                \\
                             &                &                 &                 \\
(b),BBN              &                &                 &                      \\
Best Model                   & 0.08           & 1.15            & 14.20                 \\
Conditional(1$\sigma$)       & 0.05-0.08      & 1.03-1.26       & 12.67-15.73              \\
Marginal(1$\sigma$)          &                & 0.93-1.24       &                \\
\hline
\end{tabular}
\caption{Results of the fitting analysis for $\ho=50 \kmspmpc$  
for case (a) and (b) case (i) calibration
with (BBN) and without (nBBN) BBN constraints for tilted flat CDM models.}
\label{table5}
\end{center}
\end{table}
%
%
The 68\% c.i. for the marginalized distribution of the tilt suffers a shift towards higher values of $n$ when we include the BBN constraints from 0.84-1.16 to 0.93-1.21. This is expected to be so because the BBN limits will not allow the higher values of \ob\ which seem to fit better the data.

In order to constrain $H_{0}$ for tilted CDM models consider Table~\ref{table4}.
The inclusion of BBN constraints reduces the 68\% c.i. for the marginalized distribution from $30 \kmspmpc- 70 \kmspmpc$ to $30 \kmspmpc- 55 \kmspmpc$ assuming that Saskatoon calibration is correct. Fixing $n=1$ reduces the upper limit of the \ho\ c.i. in both situations. The marginalized distribution of the Hubble constant when considered in conjunction with BBN limits gives \ho\ in the range $30 \kmspmpc - 55 \kmspmpc$ for case (i) calibration. 
\newline
\indent
In Table~\ref{table5} we also display the results obtained for 4 multipole bands for COBE data (b), assuming that Saskatoon calibration is correct and including the BBN constraints.
The best fit value of $n$ is $n=1.15 ^{+0.11}_{-0.12}$ (68 \% c.i.) while the marginalized distribution gives $n$ in the range 0.93-1.24 with 68 \% confidence.
The information contained in these 4 multipole bands seems to point for the existence of more power in the smaller scales probed by COBE and for the decrease of the normalization at low $l$. The marginal distribution suffers an insignificant variation with an interval 0.93-1.21 for case (a) and 0.93-1.24 for case (b).
These results are consistent with the inflationary prediction of $n \simeq 1$. The presence of a gravity wave component in our models would require even larger values of $n$ than those derived above. Considering Section~\ref{ten} this reaffirms our conclusion that a significant gravity wave contribution is unlikely.

\subsection{Open models}

\begin{figure}
\begin{center}
\hbox{%
\psfig{file=openI-bestmodel.ps,width=3in,angle=270}
\psfig{file=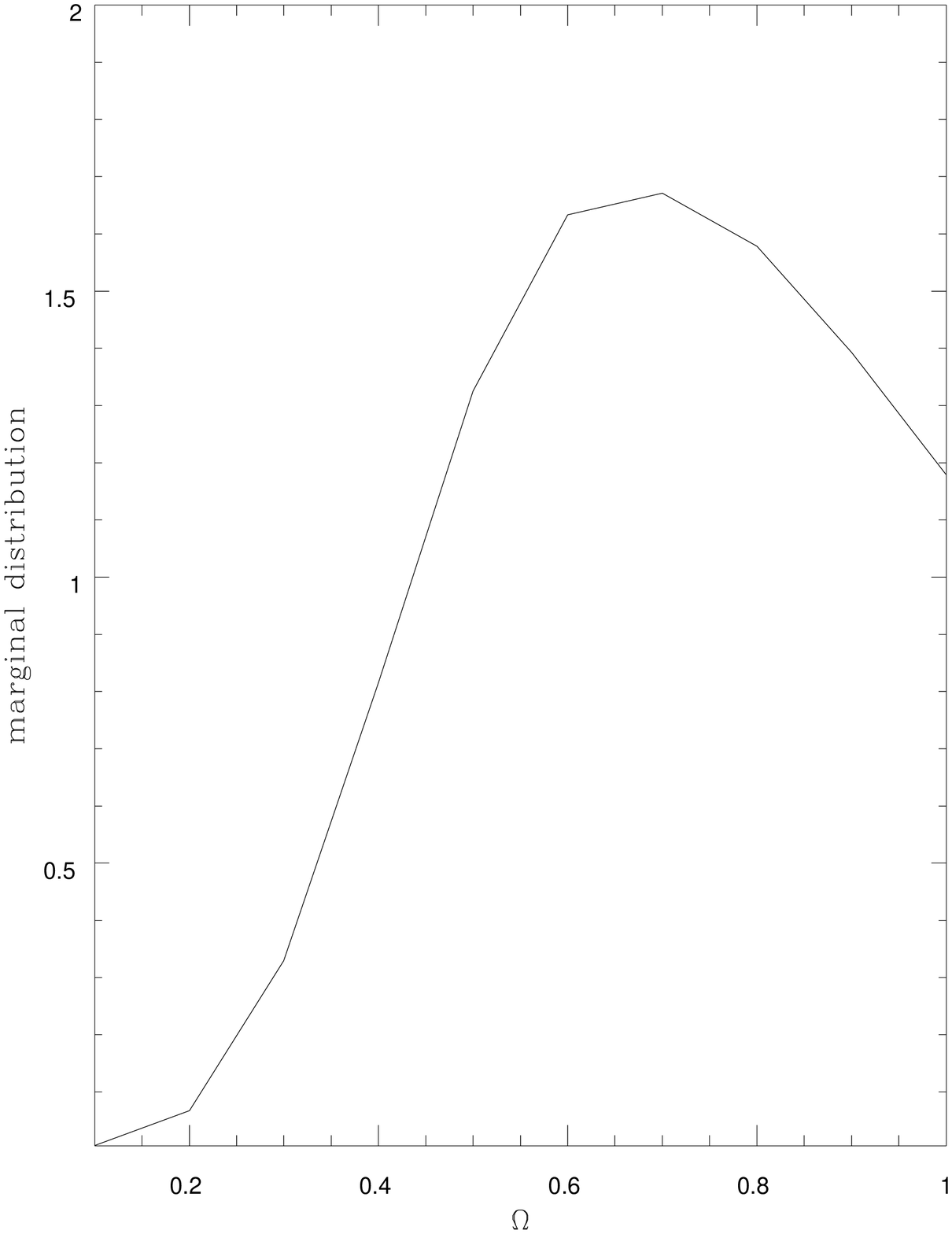,width=1.5in,height=2in}}
\hbox{%
\psfig{file=openII-bestmodel.ps,width=3in,angle=270}
\psfig{file=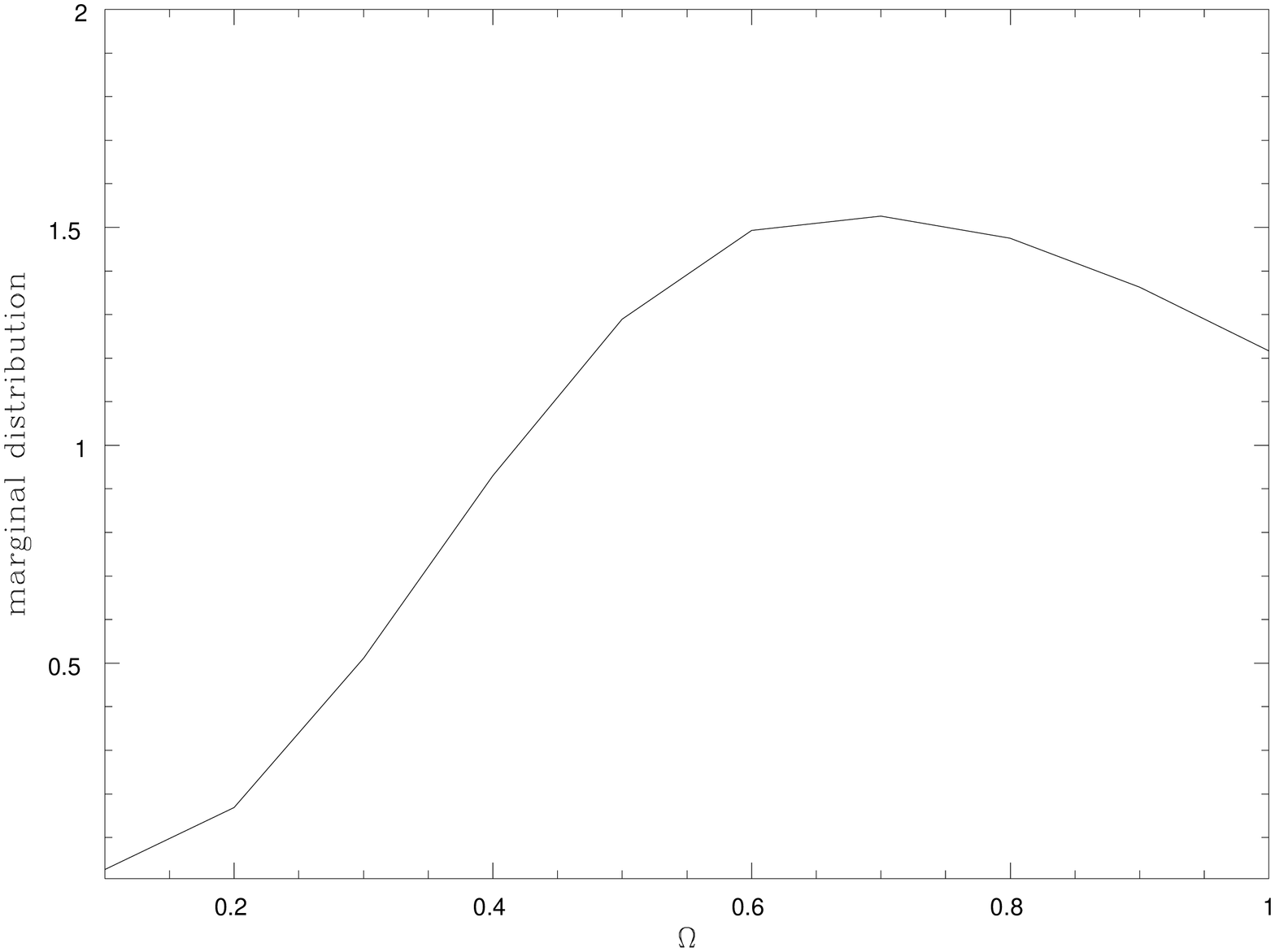,width=1.5in,height=2in,angle=270}}
\caption{For data set (a) (top figure) and for data set (c) (bottom figure): Left hand side plot shows the best-fit open model; Right hand side plot shows the marginal distribution for the parameter $\Omega$. \label{fig:openbest}}
\end{center}
\end{figure}

We initially had access to the open models ($\Omega<1$) with values for ($\Omega$,$h_{0}$) (where $h_{0}=H_{0}/(100 \kmspmpc)$) of (0.1,0.75), (0.2,0.7), (0.3,0.65), (0.4,0.65), (0.5,0.6) kindly provided by N. Sugiyama. 
Applying the analysis for case (b) ie 4 multipole bands for the COBE experiment, we conclude that $\Omega \leq 0.3$ is not compatible with the data, while the value of $\Omega =0.4$ is only allowed for the lower Saskatoon calibration data. A value of $\Omega =0.5$ offers an acceptable chi-squared fitting for both case (i) and (ii) calibrations.
This confirms the results obtained using the approximate formula of Section~\ref{comp1} reassuring us that the generalized parametrized formula  previously used constitutes a good approximation given the uncertainties in the data.
We have also considered a more complete set of open models with $\Lambda=0$ and open-bubble inflation spectrum \cite{hu,ratra2,kami}: the grid considered has $h$ values of 0.3, 0.5, 0.6, 0.7 and 0.8 (where $h=\ho/(100\kmspmpc)$), and baryon density $\Omega_b$ values of 0.01, 0.03, 0.06, together with $\Omega_b=0.0125h^{-2}$ and $0.024h^{-2}$ for each of the above values of $h$.  The $\Omega$ range considered is 0.1 to 1.0 in steps of 0.1. (These models were kindly provided by N. Sugiyama). The unavaiability of exact models for $\Omega>1$ limits some of the statistical conclusions we can draw here, but the results are still of interest. Assuming case (i) for the calibration, data set (a), and allowing the model normalisation to vary within the two sigma COBE limit we find that the best fit model has $\Omega =0.7$, $\ho=50 \kmspmpc$ and $\Omega_b=0.096$. This best fit value of $\Omega$ gives good agreement with the results obtained using the analytic approximation. Marginalizing over the other parameters, we obtain an allowed 68\% range for $\Omega$ of $0.5\le \Omega \le 1.0$. (The upper limit of 1 is due to the cutoff in the range of models considered.)
The results are presented in table~\ref{open}, assuming a similar amplitude for the the 95\% uncertainty normalization as for the flat models.
The confidence intervals for the marginal distributions of the parameters seem to be insensitive to the inclusion of new data points (as in (c)) and in particular of data from MSAM experiment. However, for data set (c), the best fit value of $H_{0}$ increases and of $\Omega_{b}$ decreases slightly, when compared with the results for case (a). For the updated data set (c) the best fit model has $\Omega =0.7$, $\ho=60 \kmspmpc$ and $\Omega_b=0.067$. 
Fig.~\ref{fig:openbest} shows the best fit open model for the fitting analysis applied to case (a) and (c) and assuming that Saskatoon calibration is correct (i) (left hand side plot);
and the marginal distribution for the parameter $\Omega$ (right hand side plot). 
\begin{table}
\begin{center}
\begin{tabular}{lccccc} \hline
                      &$\Omega$     & $H_{0}$ & $\Omega_{b}$        & $Q_{rms-ps}$        \\
(a)                   &             &         &                     &                     \\ 
Best Model              & 0.7             & 50           & 0.096          & 15.24          \\
Conditional(1$\sigma$)  & 0.5-1.0         & 50-80        & 0.05-0.096     & 13.67-16.48         \\
Marginal(1$\sigma$)     & 0.5-1.0         & 30-70        &                &                        \\
&&&& \\
(c) &&&& \\
Best Model              & 0.7             & 60           & 0.067          & 15.45         \\
Conditional(1$\sigma$)  & 0.5-1.0         & 50-80        & 0.0347-0.067   & 13.96-16.93    \\
Marginal(1$\sigma$)     & 0.5-1.0         & 30-70        &                &                  \\
(2$\sigma$)             & 0.3-1.0         & 30-80        &                &                  \\  
\hline
\end{tabular}
\caption{Results of the fitting analysis for data set (a) and (c), and case (i) calibration for open and flat models.}
\label{open}
\end{center}
\end{table}             
%

\subsection{Tilted flat \lbd\ models}

We considered, in the first instance, some of the cosmological constant, \lbd, models with  no reionization, using the Berkeley CMB code. The initially used model parameters were ($\Lambda$,$h_{0}$,$\Omega_{b}$) of (0.6,0.5,0.05), (0.6,0.5,0.03), (0.6,0.8,0.06), (0.7,0.5,0.05), (0.2,0.5,0.05).
These models seem to be consistent with the data, with large values of $\Lambda$ providing the best fit.
$\Lambda$ models compared with a flat CDM model with $h_{0}=0.45$ and $\Omega_{b}=0.1$ 
assuming case (i) calibration. 
We then considered a more complete set of flat models with a contribution from a cosmological constant, \lbd, on the top of which we add some tilt. These models angular power spectra were generated using the CMB code  provided by U.Seljac and M.Zaldarriaga. The grid considered has cosmological constant \lbd\  density \olbd\ in the range 0.3-0.7 in steps of 0.1, Hubble constant \ho in the range 30-80 $\kmspmpc$ in steps of 5 $\kmspmpc$, baryon density \ob\ in the range 0.02-0.22 in steps of 0.01 and tilt $n$ in the range 0.6-1.4 in steps of 0.1.
For details about the normalization of these models see e.g \cite{thesis}.
The fitting analysis was applied to case (a) for the COBE experiment, and case (c), case (i) calibration and case (2) for the BBN constraints. 
The results are diplayed in Table~\ref{table9}, with inclusion of the effect of tilt, with $n$ varying from 0.6 to 1.4 in steps of 0.01.
%
\begin{table}
\begin{center}
\begin{tabular}{lccccc} \hline
                      &$\Omega_{\Lambda}$     & $H_{0}$ & $\Omega_{b}$  & &           \\
(a),n=1, Qcobe& && \\ 
Best Model                & 0.7             & 55           & 0.03           & &               \\
Conditional(1$\sigma$)    & 0.4-0.7         & 55-75        & 0.03-0.06      & &               \\
Marginal(1$\sigma$)       & 0.3-0.6         & 40-65        &                & &          \\
 &&&&& \\
(a),n=1                   &                 &              &                        & $Q/Qcobe$ &        \\
Best Model                & 0.4             & 40           & 0.06                   & 0.91           &  \\
Conditional(1$\sigma$)    & 0.3-0.7         & 40-55        & 0.06-0.11              & 0.84-1.0       &  \\
Marginal(1$\sigma$)       & 0.3-0.6         & 30-60        &                        &                 &   \\
&&&&& \\
(a)                      &                 &              &                 &$n$         & $Q/Qcobe$         \\
Best Model                & 0.3             & 35           & 0.16           & 0.9        & 1.0         \\
Conditional(1$\sigma$)    & 0.3-0.7         & 30-35        & 0.11-0.16      & 0.8-1.0    & 0.91-1.1         \\
Marginal(1$\sigma$)       & 0.4-0.7         & 30-55        &                & 0.8-1.1    &                 \\
&&&&& \\
(c) &&&&& \\
Best Model                & 0.3             & 35           & 0.12           & 0.9        & 0.99            \\
Conditional(1$\sigma$)    & 0.3-0.7         & 30-40        & 0.09-0.16      & 0.8-1.0    & 0.9-1.1         \\
Marginal(1$\sigma$)       & 0.4-0.7         & 30-55        &                & 0.8-1.1    &                  \\
(2$\sigma$)               & 0.3-0.7         & 30-80        &                & 0.7-1.2    &                  \\ 
\hline
\hline
\hline
\end{tabular}
\caption{Results of the fitting analysis 
considering data set (a) and the updated data set (c), and case (i) calibration
with BBN constraints for tilted flat \lbd\ models. (Q) $Q_{rms-ps}$, (Qcobe) $Q_{rms-ps}=18$ \uk. }
\label{table9}
\end{center}
\end{table}              

\begin{figure}
\hbox{%
\psfig{figure=lambdaI-bestmodel.ps,width=3in,angle=270}
\psfig{file=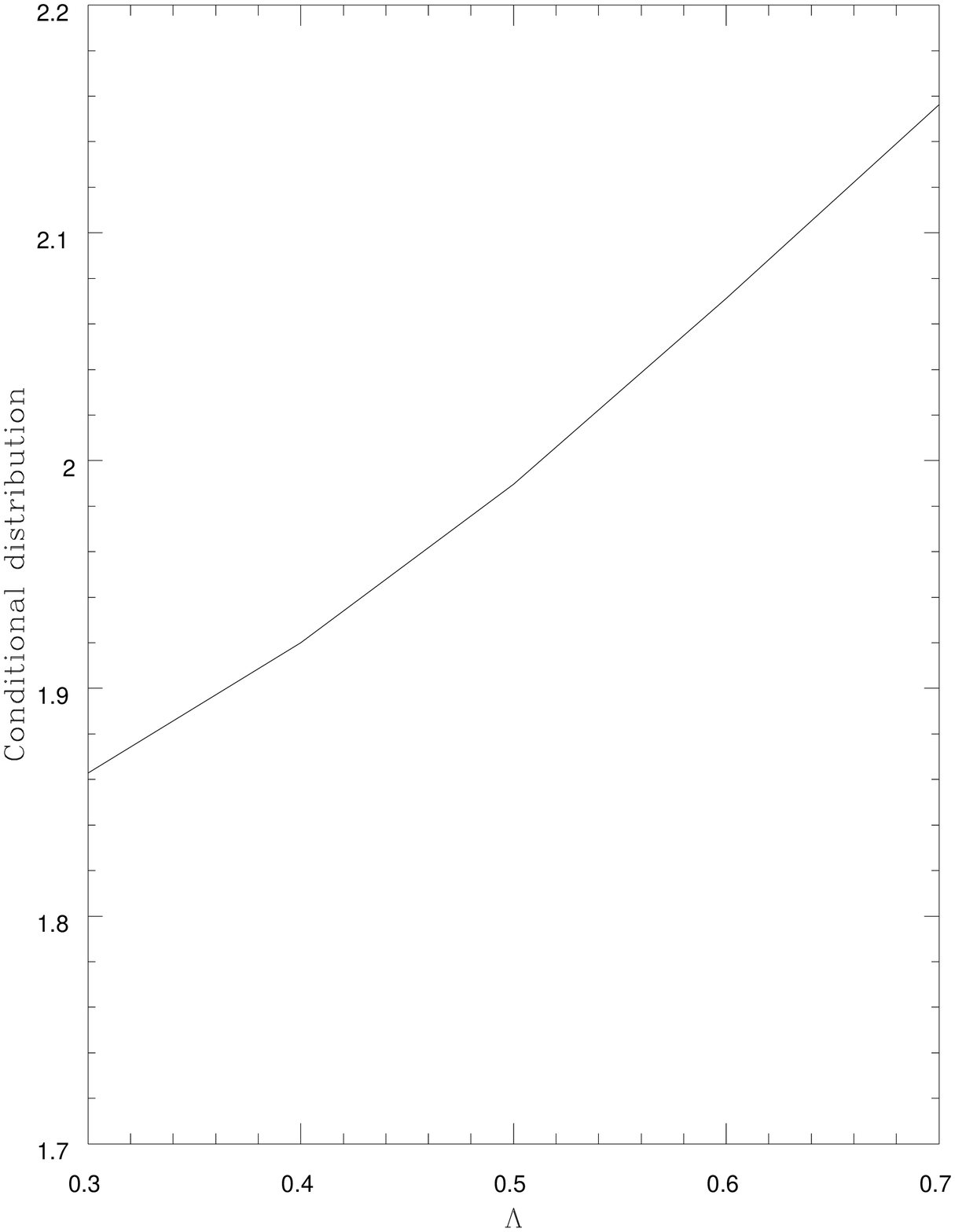,width=1.5in,height=2in}}
\hbox{%
\psfig{file=lambdaII-bestmodel.ps,width=3in,angle=270}
\psfig{file=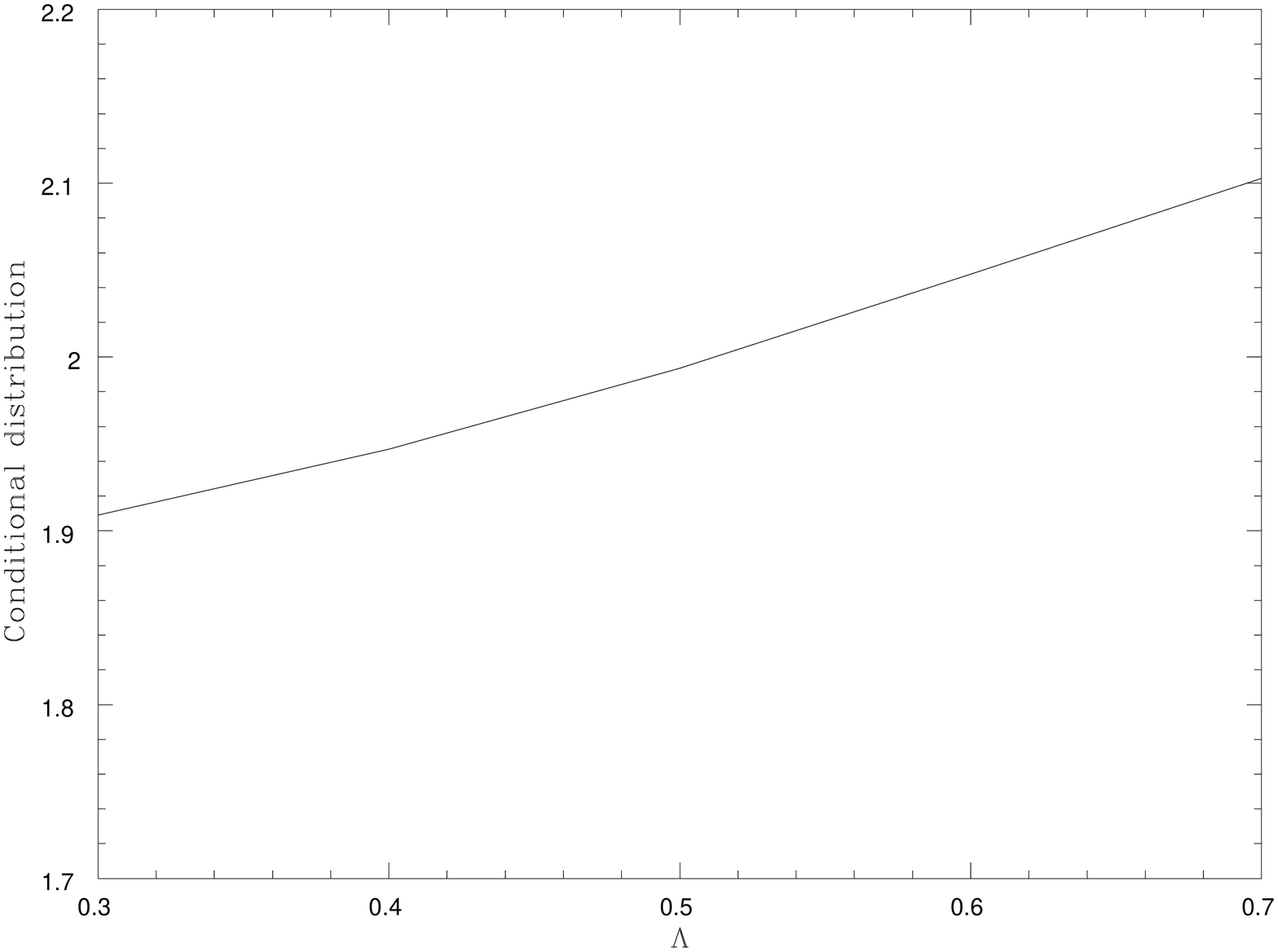,width=1.5in,height=2in,angle=270}}
\caption{For data set (a) (top figure) and for data set (c) (bottom figure): Left hand side plot shows the best-fit model for tilted \lbd\ model; results constrained to BBN limits.
Right hand side plot shows the marginal distribution for the \lbd\ parameter.}
\label{fig:lambdabest}
\end{figure}

Only fixing $n=1$ and the COBE normalization we reobtain that larger values of \lbd\ provide the best fit. The opposite is true if we include the COBE normalization uncertainty and allow the tilt to vary. In the former case the 68\% confidence interval for the marginal distribution places \lbd\ in the range 0.3-0.6, \ho\ in the range 40-65 $\kmspmpc$ while in the last case we get \lbd\ in the range 0.4-0.7, \ho\ in the range 30-55 $\kmspmpc$, tilt in the range 0.8-1.1.
These confidence intervals remain unaffected whether we consider data set (a) or data set (c). However the inclusion of new data points produces a decrease of the best fit value of $\Omega_{b}$ and $Q_{rms-ps}$. 
 The conditional distribution is not particularly discriminatory for the \lbd\ parameter and most of the models seem to offer good fit to the data. One must stress that the models considered only contemplate values of \ho and \ob\ satisfying BBN constraints. Fig.~\ref{fig:lambdabest} shows the best fit for tilted  \lbd\ models (left hand side plot) and the marginal distribution of the \lbd\ parameter (right hand side plot).  
%


\subsection{Topological defect models}

The situation for models in which structure formation is initiated by cosmic
strings \cite{strings,pen} is now more complex, since some of the predictions
for the power spectra for strings have recently changed.  

\begin{figure}
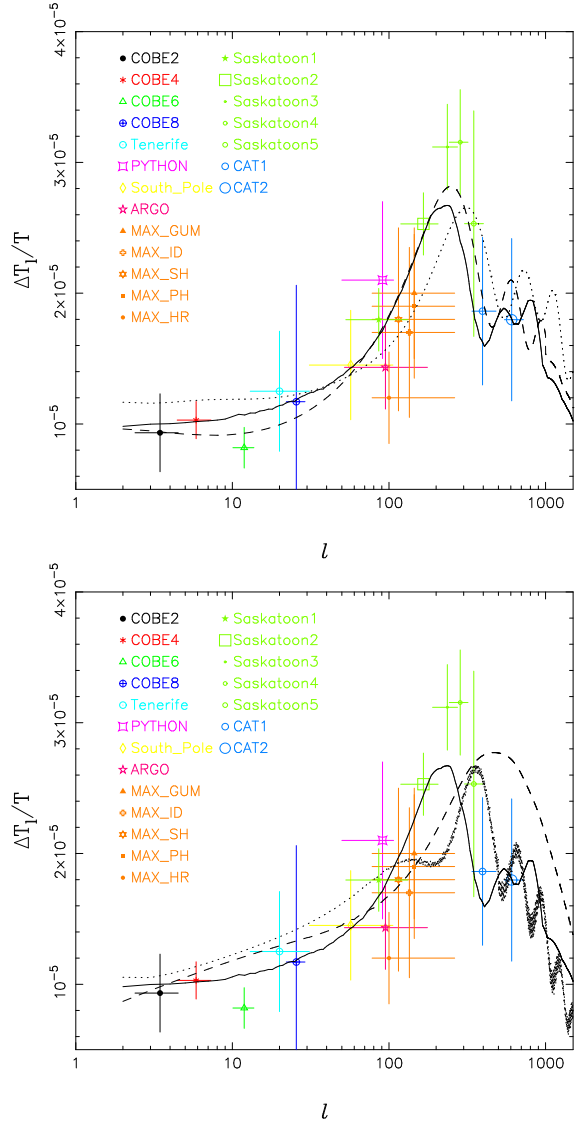

\hspace*{0.5in}
\vbox{%
\psfig{file=mod_obstegp.i.ps,width=2.95in,angle=270}
\psfig{file=mod_obstegp.d.ps,width=2.95in,angle=270}}
\caption{The data points (b) compared to the exact forms of the $C_{l}$ for on the top figure: an $\Omega=1$, $h_{0}=0.45$ and $\Omega_{b}=0.1$ standard CDM model (bold line), an $\Omega=0.5$, $h_{0}=0.6$ open model (dot line) and a flat $\Omega_{\Lambda}=0.7$, $h_{0}=0.5$ and $\Omega_{b}=0.05$ model (dashed line). For on the bottom figure: the same standard CDM model compared with a cosmic string model (dashed line) and an $\Omega=1$, $h_{0}=0.5$ and $\Omega_{b}=0.05$ texture model (dot line).
 We emphasize that the topological defect theories shown in this plot represent the initial calculations of the $C_{l}$'s and do not include the most recent predictions for the power spectrum, for these models. 
\label{fig:models1}}
\end{figure}

We initially considered a string model (J. Magueijo, priv. comm., \cite{strings} ) and texture models with ($\Omega$,$h_{0}$,$\Omega_{b}$)=(1,0.5,0.05), (1,0.7,0.05) (N. Turok, priv. comm.).
Models in which structure formation is initiated by cosmic strings \cite{strings} are consistent with the data, but are excluded for the higher calibration Saskatoon data. Texture models are also only allowed if one considers the lower calibration for the Saskatoon data. 
Previous calculations for the cosmic strings model \cite{strings} and low $\Omega$ CDM models both
have the first Doppler peak occurring in roughly the same position in $l$, so
it might be thought surprising that only the latter are eliminated by the
current data. This is traceable to the form of the low $\Omega$ power spectra
in the range $l\simeq 10$--100, where in order to match the COBE normalization
at low $l$, the models are forced to have values which are too low compared to
the data over the intermediate angular scale range. It is likely, however, that
these string models {\em will\/} be eliminated if more accurate data on the CAT
range of angular scales confirms the existing CAT results \cite{cat}.  More
recent calculations of topological defect theories indicate that the Doppler
Peak is strongly suppressed \cite{pen} and these predictions are likely to be
ruled out by the current data. 
The latest cosmic string power spectrum predictions now include a cosmological constant (Avelino \etal\ 1997, 1998, Battye \etal\ 1998). 
These predictions succeed in recovering a significant `Doppler peak' 
in the power spectrum, but have peak power for $\ell$ in the range 500 to
600, at variance with the trend of the current experimental results.
In  Fig.~\ref{fig:models1} we present the data points of data set (b) compared with a cosmic string model, the best candidate of the texture models (with $\Omega=1$, $h_{0}=0.5$ and $\Omega_{b}=0.05$) and a flat CDM model with $h_{0}=0.45$ and $\Omega_{b}=0.1$.  We emphasize that the topological defect theories shown in this plot represent the initial calculations of the $C_{l}$'s and do not include the most recent predictions for the power spectrum, for these models.
On the top plot we present the same data points compared now with candidates of both the open and $\Lambda$ models and with the same flat CDM model. 
As shown in Fig.~\ref{fig:models1} the peak of the amplitude of the angular power spectrum for these models shifts towards higher multipole $l$ (i.e. towards smaller angular scales) with respect to the standard CDM model. 
The position of the first Doppler peak for \lbd\ models is $\sim$ that of the standard CDM model as mentioned in Section~\ref{basic}.

\subsection{Conclusions}

We have presented an intercomparison of the CMB data and the theoretical fourth-year COBE normalised power spectra predicted by several models. 
For flat, scale invariant, Cold Dark Matter dominated Universe, we used our results in conjunction with Big Bang nucleosynthesis constraints to determine the value of the Hubble constant as $\ho=30-70\kmspmpc$ for baryon fraction $\Omega_b=0.02$ to $0.2$. 
We also considered a more detailed analysis of the fitting via calculation of the probability distribution function of the parameters.
We obtained for the marginalized distribution of the parameters constrained to the BBN limits, considering that Saskatoon calibration is correct, a 68\% c.i. of $30 \kmspmpc \leq \ho \leq 50 \kmspmpc$, $0.04 \leq \ob \leq 0.15$ and 16.2 \uk $\leq$ \qrms $\leq$ 19.98 \uk, confirming previous results.
Tilting the initial fluctuations away from scale invariant and fitting to the $\Delta T_{l}$ we can delimit, $n$, using the two data sets: one COBE multipole (a) and four COBE multipole bands (b) for the same range in $l$.
Assuming e.g. a CDM model without gravity waves with $\ho=50\kmspmpc$ and $\ob=0.07$ we find the primordial spectral index of the fluctuations to be $n=1.1 \pm 0.1$ (68\% c.i.), with best fit values of \qrms=15 \uk\ for $n=1.1$ (allowing for the three calibration cases).
This COBE normalised tilted model predicts a bias of the order $b \simeq 0.7$. So although this model fits well the CMB data alone it may not give a realistic scenario when we consider jointly the CMB data and the observed galaxy clustering. Aplying a more precise method to models with $\ho=50\kmspmpc$, we found for the marginalized distribution of $n$ a 68\% c.i. of $0.93\leq n \leq 1.21$ for case (a) and $0.93 \leq n \leq 1.24$ for case (b), assuming that Saskatoon calibration is correct.
Including all our set of \ho\ values we get for the marginalized distribution of the tilt a 68\% c.i. $0.85 \leq n \leq 1.18$ for case (a) and $0.80 \leq n \leq 1.15$ for the updated data set (c). 
These results are consistent with those from large scales alone given in Section~\ref{ten}, and are in close agreement with the inflationary prediction of $n \simeq 1.0$. They also state that a significant gravity wave contribution is unlikely.

Including open models in our analysis we obtain results consistent with those of Section~\ref{comp1}. For data set (a) the best fit model has $\Omega=0.7$, $\ho=50 \kmspmpc$, $\ob=0.096$ and \qrms=15.24 \uk\ while for case (c) the best fit model has $\Omega=0.7$, $\ho=60 \kmspmpc$, $\ob=0.076$ and \qrms=15.45 \uk. 
Marginalizing over the other parameters we obtain a 68\% c.i. for $\Omega$ of $0.5 \leq \Omega \leq 1.0$, where the upper limit of 1 is due to the range of models considered. 
For tilted models with non zero cosmological constant we obtain that lower values of \lbd\ provide the best fit. For data set (a) the best fit has \lbd=0.3, $\ho=35 \kmspmpc$, $\Omega_{b}=0.16$, $n=0.9$ and \qrms=18 \uk, which changes slightly for case (c) for the values of $\Omega_{b}=0.12$ and \qrms=17.82 \uk. 
The marginal distribution places \lbd\ in the range 0.4 $\leq$ \lbd $\leq$ 0.7 (68\%) and the tilt in the range  $0.8 \leq n \leq 1.1$.
In common with other recent work on fitting to current CMB data (e.g. Lineweaver \& Barbosa 1998) a tendency to low $H_0$ values (except in the case of open models) is found, although it is clear from the marginalised ranges that the statistical significance of this is not yet high. 
Work has been developed to constrain cosmological parameters using combined data from CMB and large scale structure observations. As an example, among others, see e.g. \cite {webster}, etc.
For comparison with the results of Webster \etal\ (1998), who worked jointly with recent CMB and IRAS large-scale structure results, we note that a flat $\Lambda$ model with normalization fixed to the COBE results and $n=1$ yielded a best fit of $\Omega_{\Lambda}=0.7$ and $h=0.6$.  
In conjunction with the CAT points, new results 
from the Python and QMAP experiments may soon be very significant in ruling out some models and in further delimiting the Doppler peak, sharpening up these parameter estimates. 
Recent OVRO results at $\ell \sim 590$ (Leitch, private communication) agree well with the values found by CAT. The joint CAT and OVRO results will clearly be very significant in constraining the latest cosmic string power spectrum predictions, which now include a cosmological constant \cite{pedro1,pedro2,battye}. 

For comparison we give a summary of some of the results obtained from data analysis of other experiments and in particular the results from a joint likelihood analysis of combined data sets for Open and Flat-$\Lambda$ CDM Cosmogonies.
The full SP94 data set is most consistent with $\Omega_{0} \sim 0.1-0.2$ open models and less so with old ($t_{0} \geq 15-16 Gyr$), high baryon density ($\Omega_{b} \geq 0.0175 h^{-2}$), low density ($\Omega_{0} \sim 0.2-0.4$), flat-$\Lambda$ models. The SP94 do not rule out any of the models considered at $2 \sigma$ level \cite{GRGS}. The SP94 data subsets model normalizations are mostly consistent with those deduced from the 2 yr COBE-DMR data.  
The same analysis now applied to MAX data give results consistent with those drawn from the SP94, mentioned above, but only for open models with $\Omega_{0} \sim 0.4-0.5$ the MAX and DMR normalizations overlap, the same happens for $\Omega_{0}=1$ and some higher $h$, lower $\Omega_{b}$, low-density flat-$\Lambda$ models \cite{Ganga98}.
The White Dish data does not give $2 \sigma$ detections of anisotropy for the models tested. The $2 \sigma$ upper limit for the flat bandpower angular power spectrum is 150 \uk ($Q_{rms-ps} < 96$ \uk) (HPD prescription, see \cite{GRGS} for further details). The upper limits for all models tested are consistent with those derived from the COBE-DMR data \cite{Ratra98}.
The same happens for the SuZIE data with a $2 \sigma$ upper limit for the flat bandpower angular power spectrum of
 40 \uk ($ Q_{rms-ps} < 26 $ \uk) \cite{Ganga97b}.   
The ARGO data favors an open (spatially-flat-$\Lambda$) model with $\Omega_{0}=0.23 (0.1)$. At the $2 \sigma$ confidence level model normalizations deduced from the ARGO data are consistent with those drawn from the SP94, MAX 4+5, White Dish, and SuZIE data sets. The Argo open model normalizations are also consistent with those deduced from the DMR data. However, for most spatially-flat-$\Lambda$ models the DMR normalizations are more than $2 \sigma$ above the ARGO ones \cite{Ratra99a}.
The joint likelihood analysis of SP94, ARGO, MAX, White Dish, and SuZIE data favors an $\Omega_{0} \simeq 1 (1)$ open (flat-$\Lambda$) model. Excluding SuZIE data, an $\Omega_{0} \simeq 0.3 (1)$ open (flat-$\Lambda$) model is favored.
Considering only multi-frequency data with error bars consistent with sample variance and noise considerations an $\Omega_{0} \simeq 0.1 (1)$ open (flat- $\Lambda$) model is favored. However, the data do not rule out other values of $\Omega_{0}$ in the flat-$\Lambda$ models and other values of $\Omega_{b} h^{2}$ in both models. The small-scale data normalizations are consistent at $2 \sigma$ level with those derived from the DMR data.
The Python I,II and III data combined favors an open (spatially-flat-$\Lambda$) model with $\Omega_0\simeq$ 0.2 (0.1).
At the 2 $\sigma$ confidence level model normalizations deduced from the combined Python data are mostly consistent with those drawn from the DMR, UCSB South Pole 1994, ARGO, MAX 4 and 5, White Dish, and SuZIE data sets. 
While the statistical significance of the constraints on cosmological parameters is not high, it is interesting that the combined Python data favor low-density, low $\Omega_B h^2$, young models, consistent with some of the indications from the combinations of CMB anisotropy data considered by R99b.
The $\chi^{2}$ analysis presented here favour high-density open models in variance with R99b, though in the latter case the results favour an $\Omega=1$ model when the smallest angular scale SuZIE data is included in the joint analysis. In our analysis here we include experiments such as Saskatoon and CAT which also probe higher values of $l$ when compared with the remaining observations considered. 
The difference between these two analysis might be due to the fact that they use data from different combinations of experiments. It also seems to indicate that inclusion of observations probing higher values of the multipole $l$, which do better pin down the position of the Doppler peak, shifts the density parameter to a larger value.
However, given the large error bars, both analysis are consistent for $\Omega \sim 0.4-0.5$ at $2 \sigma$ level.  


\section{Conclusions}

This paper has presented an intercomparison of CMB anisotropy experiments and models of structure formation.
On large angular scales the CMB fluctuations may probe the initial power spectrum of the fluctuations, and in particular may be used to constrain the normalization and the spectral index of an initial power-law power spectrum.
Results for both parameters were obtained for the Tenerife experiment, and an improved result achieved from the combined analysis of COBE and Tenerife data.
These experiments in conjunction with medium angular scale experiments cover angular scales ranging from $10 \dg$ to $0.25 \dg$, probing the shape of the CMB  `angular power spectrum' all way up to $l \sim 700$, offering an unique opportunity of constraining model parameters.
We carried out with a full comparison of these recent detections of CMB anisotropies in order to constrain e.g. $\Omega_{0}$, $H_{0}$, $\Omega_{b}$, $n$, 
\lbd\ etc.
A summary of the results from each of these areas is now given.

An initial intercomparison of some CDM models with `old' CMB data (available at the time the work was carried out) ruled out isocurvature models.

A summary of the analysis of the cosmic microwave background structure in the Tenerife Dec=+40\degg\ data has been presented. 
Our analysis demonstrates the existence of common structure in independent data scans at 15 and 33 GHz. For the case of fluctuations des\-cribed by a Gaussian auto-correlation function, a likelihood analysis of our combined results at 15 and 33 GHz implies an intrinsic \rms fluctuation level of $48^{+21}_{-15}$ $\mu$K on a coherence scale of $4\dg$; the equivalent analysis for a \hz model gives a power spectrum normalisation of $\qrms = 22^{+10}_{-6}$ $\mu$K. The fluctuation amplitude is seen to be consistent at the 68 \% confidence level with that reported for the COBE two-year data for primordial fluctuations described by a power law model with a spectral index in the range $1.0 \le n \le 1.6$. This limit favours the large scale CMB anisotropy being dominated by scalar fluctuations rather than tensor modes from a gravitational wave background, if power law inflation is right. 
The large scale Tenerife and COBE results are considered in conjunction with observational results from medium scale experiments in order to place improved limits on the fluctuation spectral index; we find $n=1.10 \pm 0.10$ assuming standard CDM with $\ho=50 \kmspmpc$.
For an improved grid of the parameters in conjunction with Big Bang nucleosynthesis constraints, we find $0.80 \leq n \leq 1.15$ after marginalizing the distribution function over \ho, \ob, and \qrms.
This constraint on the spectral index is in close agreement with the inflationary prediction of $n \simeq 1.0$.
   
A key prediction of cosmological theories for the origin and evolution of structure in the Universe is the existence of a `Doppler peak' in the angular power spectrum of cosmic microwave background (CMB) fluctuations. 
We presented new results from a study of recent CMB observations which provide the first strong evidence for the existence of a `Doppler Peak' localised in both angular scale and amplitude. 
This first estimate of the angular position of the peak is used to place a new direct limit on the curvature of the Universe, corresponding to a density of $\Omega=0.7^{+1.0}_{-0.4}$, consistent with a flat Universe, making use of an analytic approximate expression to describe these models.
This was further confirmed by the analysis conducted for the exact form of the angular power spectrum.  
Very low density `open' Universe models are inconsistent with this limit unless there is a significant contribution from a cosmological constant.
 Marginalizing over the other parameters, we obtain an allowed $1 \sigma$ range for $\Omega$ of $0.5\le \Omega \le 1.0$. (The upper limit of 1 is due to the cutoff in the range of models considered.)
The joint likelihood analysis carried out by R99b seems to favour low-density models in variance with the analysis carried out in Section~\ref{comp2}. The difference between these two analysis might be due to the fact that they use data from different combinations of experiments. In particular the $\chi^{2}$ analysis presented here include experiments probing a region of high multipole $l$, which do better pin down the position of the Doppler peak.
The effect of including SuZIE experiment in the analysis carried out by R99b is to shift the parameter density to a larger value, this reinforces our previous statement. However, given the large error bars, both analysis are consistent for $\Omega \sim 0.4-0.5$ at $2 \sigma$ level.
 For a flat Cold Dark Matter dominated Universe with tilt we used our results in conjunction with Big Bang nucleosynthesis constraints to determine the value of the Hubble constant as $\ho=30-55\kmspmpc$ (assuming that Saskatoon calibration is correct).
 The best fit \lbd\ model has a small $\Omega_{\Lambda}$ content, the $1 \sigma$ confidence interval for the marginal distribution places  \lbd\ in the range 0.4-0.7. 
The latest cosmic string power spectrum predictions which now include a cosmological constant might be in variance with the set of the current experimental results.

Our attempts to place increasingly more accurate limits on fundamental cosmological parameters will undoubtedly place increasing demands on observers for ever improved accuracy until the point where the intrinsic cosmic variance becomes the dominant form of error. Confidence in the results to this tolerance level will probably require close co-operation between observers and the combination of results from space-based, balloon-based and ground-based telescopes working over a range of frequencies. Potentially the most powerful observations will result from mapping overlapping/interlocking CMB fields with independent multi-frequency instruments. In contrast to the statistical results currently being reported, this method builds in the necessary redundancy to reduce systematic errors.
 As an example observations are currently in progress with the Tenerife instruments, which in conjunction with the COBE 4 year data should provide a high signal to noise map of the last scattering surface thus providing a two dimensional representation of the seed structures as compared to the simple 1-D scans reported here. Being on scales greater than the horizon size at recombination the form of such a map would reflect the structures generated in an inflationary driven phase in the very early universe.

The existence of the Doppler peak has important consequences for the future of CMB astronomy, implying that our basic theory is correct and that improving our constraints on cosmological parameters is simply a matter of improved instrumental sensitivity and ability to separate out foregrounds.  New instruments such as VSA \cite{lasenbyandhancock}, MAP and the Planck Surveyor satellite \cite{cobras} will provide this improved sensitivity and should delimit $\Omega$ and other parameters with unprecedented precision.
It is difficult using the CMB data alone to disentangle the variation due to simultaneous changes of different cosmological parameters, until such high resolution and high sensitivity maps of the sky are available.
We can however use the information from other areas of astrophysics, for example the measurements of Large-Scale Structure (LSS).
The CMB observations and the LSS measurements provide information about the power spectrum from complementary regions, with some overlap, at different cosmological times. Since these two fields provide a larger coverage of the power spectrum, it is expected that their combined use will result in more severe constraints on the cosmological parameters \cite{scott}.
In order to include overlapping sky fields and correlated data points an improvement over the  $\chi^{2}$ analysis presented here is required.
A $\chi^{2}$ analysis is no longer suitable and this must be substituted by a correlation function.
However this may require the use of supercomputers and data compression techniques since matrix sizes in the likelihood analysis will be large.

\section*{Acknowledgments} 
G. Rocha wishes to acknowledge a NSF grant EPS-9550487 with
matching support from the State of Kansas and from a K$\ast$STAR 
First award, and support from a PRAXIS XXI program of FCT (Portugal) grant. 

\end{document}